\documentclass[iop, revtex4]{emulateapj}

\usepackage{graphicx, natbib, hyperref, enumitem, amsmath}

\newcommand{\um}{$\mu$m}
\newcommand{\starsUs}{3,518,150}
\newcommand{\starsMunn}{5,216,854}
\newcommand{\stars}{8,735,004}
\newcommand{\masyear}{mas yr$^{-1}$}
\newcommand{\Hair}{\ifmmode\mskip1mu\else\kern0.08em\fi}

\slugcomment{}
\shorttitle{The MoVeRS Catalog}
\shortauthors{Theissen, West, \& Dhital}

\begin{document}

\title{Motion Verified Red Stars (M\MakeLowercase{o}V\MakeLowercase{e}RS): A Catalog of Proper Motion Selected Low-mass Stars from \textit{WISE}, SDSS, and 2MASS}

\author{Christopher A. Theissen\altaffilmark{1}, Andrew A. West\altaffilmark{1}, and Saurav Dhital\altaffilmark{2}}
\altaffiltext{1}{Department of Astronomy, Boston University, 725 Commonwealth Avenue, Boston, MA 02215, USA}
\altaffiltext{2}{Department of Physical Sciences, Embry-Riddle Aeronautical University, 600 South Clyde Morris Blvd., Daytona Beach, FL 32114, USA}

\email{ctheisse@bu.edu}

\begin{abstract}
	We present a photometric catalog of \stars\ proper motion selected low-mass stars (KML-spectral types) within the Sloan Digital Sky Survey (SDSS) footprint, from the combined SDSS Data Release 10 (DR10), Two-Micron All-Sky Survey (2MASS) Point Source Catalog (PSC), and \textit{Wide-field Infrared Survey Explorer} (\textit{WISE}) AllWISE catalog. Stars were selected using $r-i$, $i-z$, $r-z$, $z-J$, and $z-W1$ colors, and SDSS, \textit{WISE}, and 2MASS astrometry was combined to compute proper motions. The resulting \starsUs\ stars were augmented with proper motions for \starsMunn\ earlier type stars from the combined SDSS and United States Naval Observatory B1.0 catalog (USNO-B). We used SDSS+USNO-B proper motions to determine the best criteria for selecting a clean sample of stars. Only stars whose proper motions were greater than their $2$$\sigma$ uncertainty were included. Our Motion Verified Red Stars (MoVeRS) catalog is available through SDSS CasJobs and VizieR.
\end{abstract}


\section{Introduction}

	Over the past century, photometric surveys\Hair\textemdash\Hair both digital and photographic\Hair\textemdash\Hair have played an important role in many facets of astronomy. One of the largest limitations to these surveys has been object classification for point sources that have similar colors and morphologies (e.g., M giants, M dwarfs, QSOs, and distant luminous red galaxies). One method for separating nearby stellar populations from more distant objects is measuring tangential motion on the sky, or \emph{proper motion}. Proper motions are also important for distinguishing and investigating kinematically distinct populations within our Galaxy (e.g., moving groups, disk and halo stars, etc.).
	
	A number of large, all-sky catalogs of stellar positions and proper motions now exist. The United States Naval Observatory (USNO) has had a long history of tracking astrophysical objects, starting with their first published catalog UJ1.0 \citep{monet:1994:1314}, and subsequently replaced by USNO-A1.0 \citep{monet:1996:1282}, USNO-A2.0 \citep{monet:1998:120.03}, and ultimately USNO-B1.0 \citep[hereafter USNO-B]{monet:2003:984}. These catalogs are the result of the Precision Measuring Machine at the USNO Flagstaff Station, undertaking a photometric survey over $\sim$50 years on Schmidt plates. USNO-B culminated in a catalog containing the positions, proper motions, and magnitudes for over a billion objects. However, the proper motions and positions in USNO-B are relative, not absolute, making it difficult to compare observations with later epoch observations on a well-defined system \citep[e.g.,][]{roser:2008:401,roser:2010:2440}. Other proper motion catalogs of note include (but are not limited to): the SUPERBLINK catalog of northern stars with large proper motions \citep[][hereafter LSPM]{lepine:2005:1483}, the Positions and Proper Motions catalog \citep[PPM, PPMX, PPMXL;][]{roser:1993:11, roser:2008:401, roser:2010:2440}, and catalogs calibrated with the Sloan Digital Sky Survey \citep[SDSS;][]{york:2000:1579} and USNO-B \citep{munn:2004:3034, gould:2004:103} surveys.
	
	One of the newest surveys from the USNO is the CCD Astrograph Catalog \citep[UCAC;][]{zacharias:2000:2131}, now in its 4$^{\rm th}$ (and final) data release \citep[UCAC4;][]{zacharias:2013:44}. While UCAC4 does not contain as many objects as USNO-B1.0 ($\sim$10$^8$ objects), and has a shorter time baseline ($\sim$6 yrs), it is approximately five times more precise \citep{zacharias:2000:2131} due to the use of CCDs instead of photographic plates. UCAC is also in an absolute reference frame, the International Coordinate Reference System (ICRS). The newest undertaking (started in April 2012) by the USNO is the USNO Robotic Astrometric Telescope (URAT), currently in its initial data release with \citep[URAT1;][]{zacharias:2015:1}. This catalog is expected to achieve precision astrometric measurements ($\sim$10 mas) for 500 million sources, but will be relatively shallow in comparison to other surveys \citep[$R \approx 18$;][]{zacharias:2011:95}, and it will be many years before the deepest all-sky data release (northern and southern hemisphere) is available.
	
	The best example to date of high-precision space based astrometry is the survey performed by the \textit{Hipparcos} satellite \citep{1997ESASP1200.....E}, making precise astrometric measurements ($< 10$ mas) for millions of stars. The current realization of \textit{Hipparcos} data is the Tycho-2 catalog \citep{hog:2000:l27}, containing astrometric information for approximately 2.5 million stars. Prior to Tycho-2, the \textit{Hipparcos} catalog \citep{perryman:1997:l49}, containing extremely high precision astrometric measurements for 118,218 stars, defined the ICRS at optical wavelengths. The ICRS is the standard reference frame that modern surveys (e.g., SDSS, and the future Large Synoptic Survey Telescope or LSST; \citealt{ivezic:2008:}) use to calibrate their astrometry, since the majority of these surveys are tied to either Tycho-2 or UCAC. 
	
	Although many proper motion catalogs exist, they are typically tied to surveys that are biased towards the blue end of the spectrum (e.g., USNO-B). This makes most current proper motion catalogs severely incomplete at the lowest-mass end of the main-sequence, save a few smaller catalogs \citep[e.g.,][]{lepine:2005:1483, deacon:2007:163,faherty:2009:1}. It is only in recent years that an all-sky infrared point-source catalog with enough astrometric precision to build a more complete proper motion catalog for the reddest point sources has become available. Combining observations from all-sky and large area surveys taken over the past two decades will allow us to compute reliable proper motions for the lowest-mass stars.
	
	The past two decades have seen the emergence of three of the most important astronomical surveys for studies of low-mass stars: the Two Micron All-Sky Survey \citep[2MASS;][]{skrutskie:2006:1163}, SDSS, and the \textit{Wide-field Infrared Survey Explorer} \citep[\textit{WISE};][]{wright:2010:1868}. 2MASS conducted observations between 1997 and 2001 in three near-infrared (NIR) bands ($J$: 1.25 \um, $H$: 1.65 \um, and $K_s$: 2.17 \um). The 2MASS point source catalog (PSC) contains over 470 million objects. SDSS conducted visible wavelength observations starting in 2000, with some observations as recent as the last five years. SDSS observed the sky in five visible wavelength bands ($ugriz$), and Data Release 10 \citep[DR10;][]{ahn:2014:17} contains approximately 260 million point sources. \textit{WISE} began observing the entire sky in 4 mid-infrared bands (3.4, 4.6, 12, and 22 \um) starting in 2010. A second post-cryogenic mission \citep[NEOWISE;][]{mainzer:2011:53} was carried out at the end of 2010 using the two shortest bands, and surveying the entire sky over the course of a year. The AllWISE catalog combines both \textit{WISE} missions to create a catalog with enhanced photometric sensitivity and accuracy, and improved astrometric precision above each individual mission's data products. The AllWISE catalog contains over 747 million objects. Combining these three surveys provides a time baseline of $\sim$10 years.
	
	The ubiquity of M dwarfs throughout the Galaxy \citep[$\sim$70\% of the total stellar population;][]{bochanski:2010:2679}, coupled with the fact that M dwarfs have main-sequence lifetimes longer than the current age of the Universe \citep[$\sim$$10^{12}$ yrs;][]{laughlin:1997:420}, make M dwarfs important laboratories for studying numerous aspects of astronomy (e.g., Galactic and stellar evolution, kinematics, etc.). Recent results have also suggested that M dwarfs have a strong penchant for building terrestrial planets \citep[e.g.,][]{dressing:2013:95, dressing:2015:45}. This affinity for creating terrestrial planets, coupled with the relative ease for finding terrestrial planets around M dwarfs (due to their size ratios and small orbital distances), make M dwarfs important hosts for studying Earth-sized planets and habitability throughout the Galaxy. 
		
	Many large catalogs of M dwarfs (dMs) currently exist, including the Palomar/Michigan State University \citep[PMSU;][]{reid:1995:1838, hawley:1996:2799} survey ($\sim$2400 spectroscopic dMs), and the SDSS Data Release 7 \citep[DR7;][]{abazajian:2009:543} spectroscopic catalog \citep[70,841 spectroscopic dMs;][]{west:2011:97}. However, such catalogs make up only a small fraction of the millions of photometric dMs contained within SDSS. For example, \citet{bochanski:2010:2679} used SDSS DR7 to retrieve $\sim$15 million photometrically selected, but not proper motion verified, dMs to investigate the mass and luminosity functions of the Galactic disk. While most of the red point sources in SDSS are M dwarfs rather than giants or red galaxies for the color and magnitude range chosen by \citet{bochanski:2010:2679}, proper motions can help to select \textit{bona-fide} low-mass stars.
	
	\textit{Gaia} \citep{perryman:2001:339} is currently conducting the largest astrometric survey to date. \textit{Gaia} is a magnitude limited survey, the limits of which are shallower \citep[$r \leq 20$;][]{ivezic:2012:251} than the combined \textit{WISE}+SDSS limits ($r \leq 22.2$), making \textit{Gaia} incomplete for the faintest and lowest-mass (reddest) stars. Approximately 60\% of the stars in the combined photometric dataset of \textit{WISE}, SDSS, and 2MASS have $r\geq20$, making the majority of low-mass stellar candidates beyond the reach of \textit{Gaia} \citep[for relative point-source densities at different $r$ magnitudes see Figure 6 of][]{bochanski:2010:2679}.
	
	To make use of a larger photometric sample of dMs within SDSS, we combine \textit{WISE}, SDSS, and 2MASS observations to compute proper motions over $\sim$10 year baselines for photometrically selected objects with dM colors. After a color cut and selection of reliable proper motions, we are left with a sample of \stars\ stars. In Section~\ref{methods}, we outline our methods for computing proper motions and errors. We also estimate the intrinsic uncertainty within our catalog using SDSS selected quasars. In Section~\ref{data}, we discuss the selection criteria used to build our photometric dM sample, and address contamination (Section~\ref{contam}). In, Section~\ref{reliability} we assess the reliability of our catalog. We augment our derived proper motions with measurements from SDSS+USNO-B in Section~\ref{usno}. In Section~\ref{movers}, we discuss the properties of our Motion Verified Red Stars (MoVeRS) catalog and how to query it. We describe preliminary science results that can be achieved with this catalog in Section~\ref{discussion}. Our summary follows in Section~\ref{summary}.

\section{Methods: Combining \textit{WISE}, SDSS, and 2MASS}\label{methods}

\subsection{Astrometric Algorithms}\label{pmalg}
	
	SDSS was originally calibrated against UCAC and Tycho-2 \citep{pier:2003:1559}, but as of its seventh data release (DR7) was calibrated against UCAC2 and an internal UCAC release known as ``r14\footnote{\url{https://www.sdss3.org/dr10/algorithms/astrometry.php}}." This recalibration reduced systematic errors from $\sim$75 mas (Tycho-2) and $\sim$45 mas (UCAC) to less than 20 mas\footnote{\url{http://classic.sdss.org/dr7/algorithms/astrometry.html}}. SDSS is also on the ICRS since the Tycho-2 catalog is based upon \textit{Hipparcos} astrometry, which defines the ICRS at visible wavelengths. All SDSS astrometric calibrations for this study were performed in the $r$-band.
	
	We noticed that SDSS positional errors (\textsc{raErr}, \textsc{decErr}) in all SDSS Data Releases more recent than DR7 are pixel-centroiding errors rather than absolute astrometric errors (B. A. Weaver, personal communication). To compute absolute astrometric errors, we found the total error (centroiding plus calibration) in great circle coordinates using the following equations,
\begin{equation}
\sigma_\mu = \sqrt{ (\textsc{rowcErr} \times 0.3961)^2 + \textsc{muErr}^2}
\end{equation}
and
\begin{equation}
\sigma_\nu = \sqrt{ (\textsc{colcErr} \times 0.3961)^2 + \textsc{nuErr}^2},
\end{equation}
where \textsc{rowcErr} and \textsc{colcErr} are the row center and column enter position errors in $r-$band coordinates, respectively, and are found in the SDSS CasJobs ``\textsc{PhotoObj}" table. The fields \textsc{muErr} and \textsc{nuErr} are the astrometric errors in Great Circle coordinates ($\mu$ and $\nu$) for the $r$-band, and are found in the ``Field" table. The factor of 0.3961 is the SDSS pixel scale (arcsec pix$^{-1}$). Next, using the above total errors in Great Circle coordinates, we converted to $\alpha$ and $\delta$ through the following equations, 
\begin {equation}
	\begin{aligned}
	s = -\sin(\textsc{incl}) \sin(\textsc{nu}) &\sin(\textsc{mu}-\textsc{node}) \\
	&+ \cos(\textsc{incl}) \cos(\textsc{nu}),
	\end{aligned}
\end{equation}
\begin {equation}
	c = -\sin(\textsc{incl}) \cos(\textsc{mu}-\textsc{node}),
\end{equation}
\begin {equation}
	\sigma_\mathrm{\alpha_\mathrm{SDSS}} = \sqrt{(c\cdot \sigma_\mu)^2 + (s \cdot \sigma_\nu)^2},
\end{equation}
and
\begin {equation}
	\sigma_\mathrm{\delta_\mathrm{SDSS}} = \sqrt{(s \cdot \sigma_\mu)^2 + (c \cdot \sigma_\nu)^2},
\end{equation}
where \textsc{mu}, \textsc{nu}, \textsc{node}, and \textsc{incl} refer to fields in the SDSS CasJobs ``Frame" table. \textsc{mu} and \textsc{nu} refer to the Great Circle coordinates of the frame center, and \textsc{incl} and \textsc{node} are the inclination and right ascension of the ascending node of the scan Great Circle with respect to the J2000 celestial equator\footnote{\url{https://www.sdss3.org/dr10/algorithms/astrometry.php}}.

	The 2MASS PSC uses the Tycho-2 catalog to reconstruct its coordinates in the ICRS\footnote{\url{http://www.ipac.caltech.edu/2mass/releases/allsky/doc/sec2_2.html}}, with accuracies between 70--120 mas. 2MASS astrometric errors are reported as an error ellipse with the entries \textsc{err\_maj} ($\sigma_\mathrm{MAJ}$), \textsc{err\_min} ($\sigma_\mathrm{MIN}$), and \textsc{err\_ang} ($\sigma_\mathrm{\theta}$). These were converted to $\sigma_\alpha$ and $\sigma_\delta$ components using the following equations,
\begin{equation}
	\sigma_{\alpha_\mathrm{2MASS}} = \sqrt{(\sigma_\mathrm{MAJ} \cdot \sin\sigma_\theta)^2 + (\sigma_\mathrm{MIN} \cdot \cos\sigma_\theta)^2}
\end{equation}
and
\begin{equation}
	\sigma_{\delta_\mathrm{2MASS}} = \sqrt{(\sigma_\mathrm{MAJ} \cdot \cos\sigma_\theta)^2 + (\sigma_\mathrm{MIN} \cdot \sin\sigma_\theta)^2}.
\end{equation}
	Relative astrometric calculations for 2MASS and \textit{WISE} are computed on the unit sphere, and therefore the $\alpha$ component of the astrometric uncertainties already accounts for the $\cos\delta$ term (Vandana Desai, personal communication).
	
	\textit{WISE} is tied to the ICRS through 2MASS. However, to address possible systematic proper motion shifts between the two catalogs due to their different epochs, the \textit{WISE} pipeline used proper motion data from UCAC4 to readjust 2MASS positions before they were used as reference stars\footnote{\url{http://wise2.ipac.caltech.edu/docs/release/allwise/expsup/sec2_5.html}}. Estimated errors for the source catalog are $< 100$ mas. \textit{WISE} astrometric errors are denoted by the entries \textsc{sigRA} ($\sigma_\textsc{sigRA}$) and \textsc{sigDEC} ($\sigma_\textsc{sigDEC}$).	
	
	For sources with only two observation epochs, the uncertainties for each proper motion component are given as
\begin{equation}
	\begin{aligned}
	\sigma_{\mu_\alpha}^2 = \left(\frac{\cos \bar{\delta}}{\Delta_t}\right)^2 \bigg[\sigma_{\alpha_1}^2 + \sigma_{\alpha_2}^2 &+ \left(\frac{\Delta_\alpha}{\Delta_t}\right)^2 (\sigma_{t_1}^2 + \sigma_{t_2}^2) \\
	&+ \sigma_{\bar{\delta}}^2 \tan^2 \bar{\delta} \bigg] 
	\end{aligned}
\end{equation}
and
\begin{equation}
	\sigma_{\mu_\delta}^2 = \Delta_t^{-2} \bigg[\sigma_{\delta_1}^2 + \sigma_{\delta_2}^2 + \left(\frac{\Delta_\delta}{\Delta_t}\right)^2 (\sigma_{t_1}^2 + \sigma_{t_2}^2)\bigg],
\end{equation}
where $\Delta_\alpha = \alpha_2-\alpha_1$, $\Delta_\delta = \delta_2-\delta_1$, $\Delta_t = t_2-t_1$, $\bar{\delta}$ is the weighted-mean declination, and $\sigma_{\bar{\delta}}$ is the error in the weighted mean declination. Here, $\alpha$, $\delta$ and $t$ refer to the position and time, with ``1" being the first epoch and ``2" the second epoch. The final term in $\sigma_{\mu_\alpha}$ is orders of magnitude smaller than 1 \masyear\ and is therefore negligible. Proper motion errors for the $\alpha$ component are in proper units (i.e. $\Delta_\alpha \cdot \cos \delta$). 

\begin{figure}
\centering
 \includegraphics{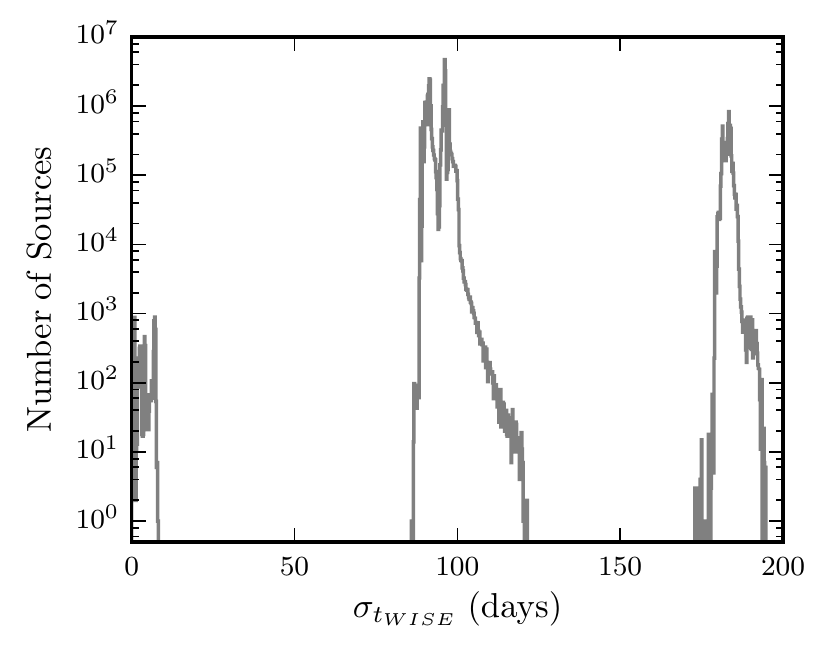}
\caption{\textit{WISE} temporal uncertainties for our sources. The large uncertainties are due to the AllWISE catalog being a composite of \textit{WISE} and \textit{NEOWISE} observations. This temporal uncertainty will account for $<$ 1 \masyear\ in our astrometric solution for stars with baselines $> 4$ years, however, we account for it since a small subset of our observations have baselines of $\sim$1 year.
\label{fig:wisetimesig}}
\end{figure}

	On average, the temporal uncertainty is much smaller than the time baseline between measurements ($\sigma_t \approx 60$ seconds for 2MASS and SDSS). The AllWISE catalog combines observations from the initial \textit{WISE} mission and the post-cryogenic \textit{NEOWISE} survey. We define the temporal uncertainty to be halfway between the difference of the most recent observation and the earliest observation in the $W1$-band (i.e. $\sigma_{t_\mathrm{WISE}} =$ [\textsc{w1mjdmax} - \textsc{w1mjdmin}]/2). As is shown in Figure~\ref{fig:wisetimesig}, this causes some observations to have temporal uncertainties between 80--200 days. For sources with time baselines of at least four years, this uncertainty is $\lesssim 1$ \masyear. However, a small fraction of our sources have time baselines of about one year. For this reason, we do not remove the temporal uncertainty term of the proper motion error. Our motivation for using the AllWISE catalog rather than the All-Sky Release Catalog is that the AllWISE catalog has better astrometric accuracy. This is due to the inclusion of proper motions to correct 2MASS astrometric reference stars for the greater than nine year baseline between the \textit{WISE} and 2MASS surveys.

	For sources with three observational epochs (i.e. \textit{WISE}, SDSS, and 2MASS), we computed a weighted linear fit to the positions versus time. Rather than use an linear least squares approach, which requires the uncertainty in the independent variable (in this case time), to be negligible, we chose to invoke an Orthogonal Distance Regression \citep{boggs:1990:183} method to calculate proper motions in each component ($\alpha, \delta$) separately. This allows us to take into account the sometimes significant temporal uncertainties on \textit{WISE} observations.

\subsection{Precision: Measuring the Motions of SDSS Quasars}\label{quasars}
	
\begin{figure}
\centering
 \includegraphics{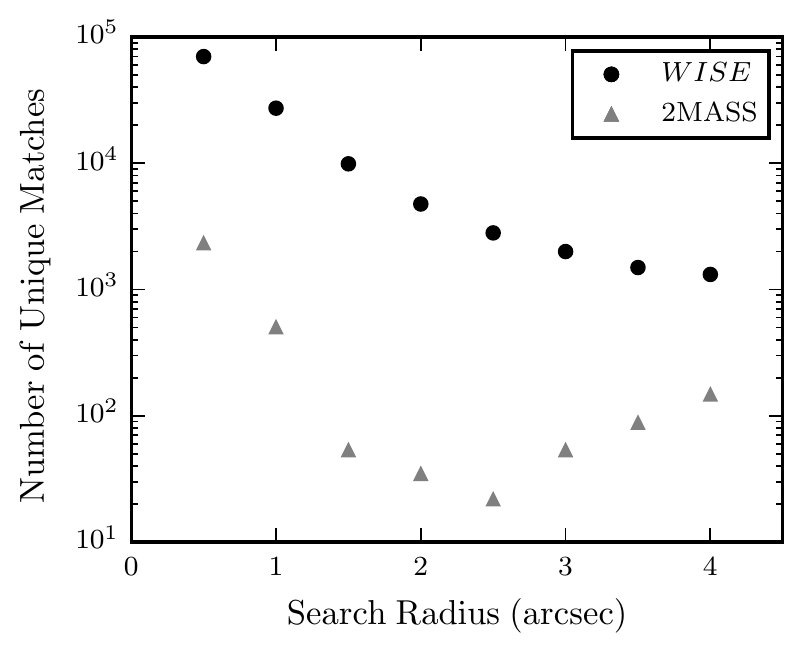}
\caption{Unique SDSS QSO matches per search radius in 2MASS and \textit{WISE}. Both reach a maximum at a search radius of 0.5\arcsec. The increasing number counts at $>$ 2\arcsec\ for 2MASS indicate resolved neighboring objects being pulled into our search radius for faint QSOs that 2MASS is not able to detect.
\label{fig:QSO_match}}
\end{figure}

	To investigate the intrinsic error in the proper motions for each survey, we required stable objects on the sky with essentially zero tangential motion. Quasars make ideal calibrators due to their high luminosities and extragalactic distances. We used the DR10 SDSS quasar catalog \citep{paris:2014:a54}, which contains 166,583 spectroscopically confirmed quasars. Using this catalog, we cross-matched to both the 2MASS PSC and \textit{WISE} AllWISE source catalog. Matching was done using search radii in steps of 0.5\arcsec\ out to 4\arcsec, the results of which are shown in Figure~\ref{fig:QSO_match}. The number of unique matches reaches a maximum at 0.5\arcsec\ for both 2MASS and \textit{WISE} matches. Since we sought to estimate the precision of our catalog, we chose to only use matches within a search radius of 0.5\arcsec. This left us with 69,949 matches between SDSS and \textit{WISE} and 2351 matches between SDSS and 2MASS, 2283 of which were common in both \textit{WISE} and 2MASS. The low number of matches between SDSS and 2MASS is due to the difference in wavelengths and relative depths between the two surveys. The magnitude and flux limits of each survey are shown in Table~\ref{tbl:limits}. Because we were assessing the intrinsic precision of the catalog, we chose to apply further cuts to retain only the most pristine detections of QSOs for our analysis. The following criteria were required:
	
\begin{deluxetable}{l c c c}
\tabletypesize{\scriptsize}
\tablecolumns{4}
\tablecaption{Limits for \textit{WISE}, SDSS, and 2MASS\label{tbl:limits}}
\tablehead{
\colhead{Survey}	& \colhead{Band} 	& \colhead{Limiting Magnitude}		& \colhead{Limiting Flux}\\
				&				& \colhead{(mags)}				& \colhead{(ergs s$^{-1}$ cm$^{-2}$ $\textup{\AA}^{-1}$)}
}
\startdata
SDSS		& $r$		& 22.2 (AB)	& $\sim$$4\times10^{-18}$\\
SDSS		& $i$ 		& 21.3 (AB)	& $\sim$$6\times10^{-18}$\\
SDSS		& $z$		& 20.5 (AB)	& $\sim$$8\times10^{-18}$\\
2MASS		& $J$ 		& 15.8 (Vega)	& $\sim$$2\times10^{-17}$\\
2MASS		& $H$	 	& 15.1 (Vega)	& $\sim$$4\times10^{-17}$\\
2MASS		& $K_s$		& 14.1 (Vega)	& $\sim$$8\times10^{-17}$\\
\textit{WISE}	& $W1$		& 17.1 (Vega)	& $\sim$$10^{-18}$
\enddata
\end{deluxetable}

\begin{enumerate}[noitemsep]
\item SDSS \textsc{clean} $=1$,
\item 2MASS \textsc{cc\_flg} $=$ 000,
\item \textit{WISE} \textsc{cc\_flag} $=$ 0000,
\item \textit{WISE} \textsc{w1snr} $>= 30$,
\item 2MASS \textsc{j\_snr}, \textsc{h\_snr}, or \textsc{k\_snr} $>= 10$,
\item 2MASS \textsc{gal\_contam} $=$ 0,
\item \textit{WISE} \textsc{ext\_flag} $=$ 0,
\item $\sigma_\alpha < 175$ mas \& $\sigma_\delta < 175$ mas, and
\item The closest neighboring primary SDSS object was greater than 6.1\arcsec\ (the $W1$ FWHM) from our source.
\end{enumerate}
These cuts left us with 447 matches for sources with \textit{WISE}+SDSS+2MASS, 4091 matches for sources with only \textit{WISE}+SDSS, and 413 matches for sources with only SDSS+2MASS.

\begin{figure*}
\centering
 \includegraphics{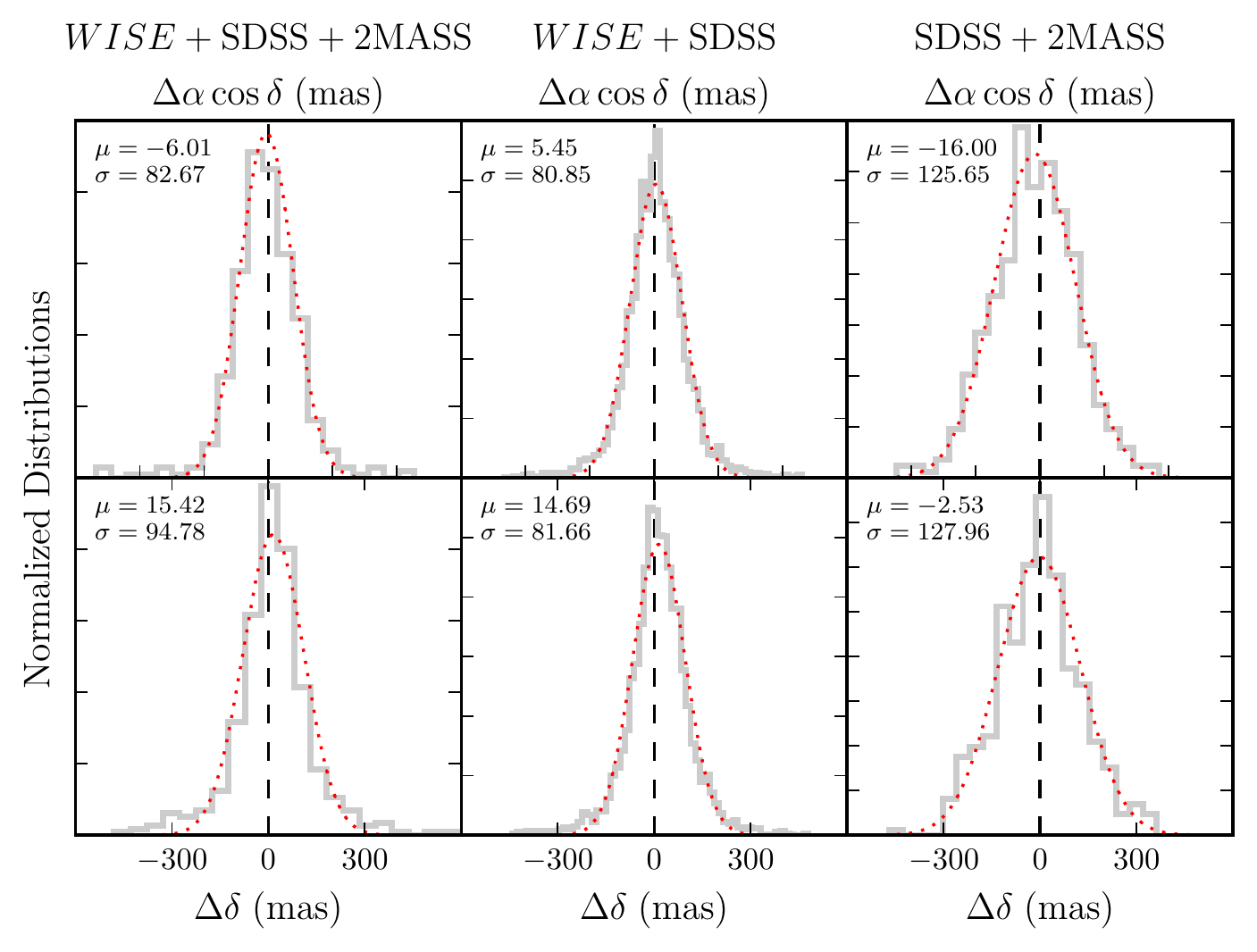}
\caption{Normalized distributions for the positional difference components among each of the surveys (gray). The best-fit normal distribution is also plotted (red dotted line). The precision tends to be better for observations with a \textit{WISE} epoch. These uncertainties are weighted by the time baseline between observations and added in quadrature to the fitting errors (Section~\ref{pmalg}).
\label{fig:QSOdist}}
\end{figure*}

	We applied the above algorithm to compute the angular distance measured from the surveys. Distributions of our computed angular distances are shown in Figure~\ref{fig:QSOdist}. The estimated errors in our sample of QSOs are in agreement with, or slightly better than, reported positional uncertainties among each of the three surveys. The largest uncertainties are in our SDSS+2MASS baseline, however, these objects make up a small fraction of our entire proper motion catalog (see Section~\ref{data}). Figure~\ref{fig:QSOdist} represents the intrinsic positional errors in each of our fits ($\sim$90 mas for \textit{WISE}+SDSS+2MASS, $\sim$80 mas for \textit{WISE}+SDSS, and $\sim$125 mas for SDSS+2MASS). We add this error, weighted by the time baseline, to the proper motion error for each component.
	
	A number of studies have attempted to correct systematic errors in proper motion catalogs using QSOs \citep[e.g.,][]{roser:2010:2440,wu:2011:1313,lopez-corredoira:2014:a128,grabowski:2015:849}. We did not observe any large systematics within the \textit{WISE}+SDSS sources, and the small sample size among all other surveys did not allow us to investigate or correct for any potential systematics offsets. 
	
\section{Data}\label{data}

\subsection{Building the Catalog}\label{catalog}

	SDSS DR10 boasts over 900 million unique optical sources (after accounting for multiple epochs of photometry)\footnote{\url{https://www.sdss3.org/dr10/scope.php}}. This study focused on the low-mass stars (point sources) within DR10. We used the following photometric selection criteria, many of which were adapted from \citet[hereafter B10]{bochanski:2010:2679}, for our initial search:
	
\begin{enumerate}
\item The photometric objects were flagged as \textsc{primary}. This was done by querying the \textsc{Star} sub-catalog of the \textsc{PhotoPrimary} catalog. This also ensured a morphological classification of a point source.
\item The photometric objects fell within the following color limits:
$$16 < r < 22,$$
$$i < 22,$$
$$z < 21.2,$$
$$r-i \geqslant 0.3, \mathrm{and}$$
$$i-z \geqslant 0.2.$$
These color criteria allow for sources slightly bluer than typical dM colors, but were chosen to be inclusive for this stage of the selection process. The $i$- and $z$-band magnitude limits extend past the SDSS 95\% completeness ($i < 21.3$ and $z < 20.5$) limits, but we apply additional cuts below. The $r$-band limit removes saturated sources and is slightly brighter than the 95\% completeness limit ($r < 22.2$). Although we obtain sources past the 95\% completeness limit, we apply more stringent criteria later.
\item We removed sources that had a flag in the $r$-, $i$-, or $z$-bands indicating: 
\begin{enumerate}
\item saturated photometry (\textsc{saturated}); 
\item a significant amount ($>$20\%) of the flux was interpolated from the point spread function (PSF; \textsc{psf\_flux\_interpreted})
\item centroiding failure caused center to be determined by peak pixel (\textsc{peakcenter});
\item too few good pixels for an interpolated source, causing errors to be underestimated (\textsc{bad\_counts\_error}); 
\item center pixel was too close to interpolated pixel (\textsc{interp\_center}), and included pixels that were not checked for peaks, potentially saturated (\textsc{notchecked}); 
\item after deblending the object did not have a peak (\textsc{deblend\_nopeak}); and 
\item contained a pixel interpreted to be part of a cosmic ray (\textsc{cosmic\_ray}).
\end{enumerate}
\item Sources that indicated centroid was not determined from the $r$-band, but were transformed from some other band (\textsc{canonical\_center}), were removed. This was done to ensure objects with reliable $r$-band astrometry, which was used as the calibrator for our SDSS baseline.
\end{enumerate}
These cuts ensured that the $riz$ photometry for each source were reliable. Applying these criteria returned 69,792,454 objects.
	
	Using a 6\arcsec\ search radius (a similar matching radius to other searches for ultra-cool dwarfs, e.g., \citealt{zhang:2009:619}), we matched the SDSS photometric stars to \textit{WISE} and 2MASS sources requiring:
\begin{enumerate}
\item 2MASS \textsc{j\_psfchi}, \textsc{h\_psfchi}, or \textsc{k\_psfchi} $\leq 2$. This ensures a point-like source morphology, and increases our confidence for a good astrometric measurement.
\item \textit{WISE} \textsc{w1rchi2} $\leq 3$. This is the same requirement used to determine a single-point source from a blended object (after deblending) by the \textit{WISE} photometric pipeline.
\item \textit{WISE} \textsc{SNR$_{W1}$} $>0$. This removes sources that were not detected in $W1$, but detected in a longer wavelength band. The spectral energy distributions (SEDs) of low-mass stars peak in the near-infrared, therefore, as $W1$ was the deepest of the \textit{WISE} bands, low-mass stars should have a detection in $W1$ if they have a detection at a longer wavelength. 
\end{enumerate}
After this cut we were left with 20,164,221 matches with entries in all three catalogs, 22,741,703 with only \textit{WISE}+SDSS detections, and 2,947,606 with only SDSS+2MASS detections.

\subsection{Tracing the Stellar Locus}\label{ms}

	We were interested in selecting point sources that have the expected colors of main-sequence stars for our initial sample. To ensure this, we computed the expected main-sequence colors for low-mass and very-low-mass stars in the \textit{WISE}, SDSS, and 2MASS photometric systems. We required the following criteria, adapted from \citet{davenport:2014:3430} for low-mass stars, and the AllWISE Explanatory Supplement\footnote{\url{http://wise2.ipac.caltech.edu/docs/release/allwise/expsup/sec2_4a.html}}, to select high quality matches:
\begin{enumerate}
\item $13.8 < r < 21.5$ \& $\sigma_{r,i,z} < 0.05$, this cut ensured precision photometry within the SDSS 95\% completeness limit and saturation limit, and ensured good morphological classification, which has been shown to have an error rate of 5\% at $r=21$ \citep{lupton:2001:269}.
\item $J > 12$ \& $\sigma_{J} < 0.05$, this cut ensured precision photometry, and should have removed giant stars by using a lower limit on the magnitude \citep{covey:2008:1778}.
\item $W1 < 17.1$ \& $\sigma_{W1} < 0.05$, this cut also ensured precision photometry and selected only sources within the AllWISE 95\% completeness limit.
\item $|b| > 20^\circ$, this cut reduced extinction effects by removing candidates near the Galactic plane.
\end{enumerate}
After applying the aforementioned cuts, we were left with 9,298,344 sources with only SDSS photometry, 9,890,521 sources with \textit{WISE}+SDSS photometry, and 2,390,962 sources with SDSS+2MASS photometry.

	Previous studies have suggested that the majority of point-like sources that meet our color selection criteria (Section~\ref{catalog}) are stars versus distant Galaxies \citep{bochanski:2010:2679}, however, we wish to be more selective, choosing only high probability stellar candidates for our catalog. To do this, we took our above candidates, selected for their reliable photometry, and examined the stellar locus in bands from all three surveys. Many previous studies have investigated the stellar color locus in numerous colors and a number of photometric systems \citep[see][and references therein]{davenport:2014:3430}. Instead of using previous results, we chose to measure the color locus from our stellar sample chosen above. We also chose to only measure the locus for the reddest SDSS bands ($riz$) and the deepest 2MASS and \textit{WISE} bands ($J$ and $W1$, respectively). Our chosen colors for computing the stellar locus were $r-i$ versus $i-z$, $r-z$ versus $z-J$, and $r-z$ versus $z-W1$.

	In small steps of $\delta$(color), we computed the median absolute deviation from the median and removed objects greater than 5 times the median of the deviations until our distribution converged (i.e. there were no further sources to remove). We then computed the mean and standard deviation of the remaining color distribution. We used a bin size of 0.01 mags for the high density areas (the middle of the distribution), taking steps of 0.01 mags and recomputing. For the red end of the distributions (the low density areas), we increased the bin size to 0.2 mags. Averages and 1$\sigma$ colors, color steps, bin sizes, and the number of stars in each bin can be found in Appendix~\ref{polys}. The visual representation of our computed means and 3$\sigma$ deviations are shown in Figure~\ref{fig:ColorCuts}, where it can be seen that we trace the source density, removing large color outliers.
	
	Due to our initial selection criterion for sources along low-extinction sight-lines ($|b|>20^\circ$), it is possible that our sample could be biased towards stars with low extinction values. Rather than apply an extinction correction to our stars, we investigated how extinction may bias our selection criteria. To test how accurate our color criteria are for selecting low-mass stars and the effects of interstellar reddening, we applied our cuts to the SDSS DR7 spectroscopic sample of 70,841 M dwarfs \citep[][hereafter W11]{west:2011:97}, all of which have estimated extinction values from \citet{jones:2011:44}. We investigated the fraction of returned W11 stars, after passing them through our color selection criteria. We applied the cuts above and the cuts from Section~\ref{catalog} to the W11 catalog, excluding the requirement that stars have $|b| > 20^\circ$ since we were interested to see how extinction affected our selection method. The total return fraction for stars with $A_V \leq 0.5$ and $A_V > 0.5$ are shown in Table~\ref{tbl:matches} for each combination of color selection criteria. For stars with $A_V \leq 0.5$, our color selection criteria returned $\geq$ 95\% of the W11 inputs stars for all color selection criteria. Even for stars with $A_V > 0.5$ our color selection criteria returned more than 92\% of the W11 input stars.
	
	The W11 catalog also contains both disk dwarfs and subdwarfs \citep{savcheva:2014:145}, populations that can be separated by their metallicity using the $\zeta$-spectroscopic index \citep[e.g.,][]{dhital:2012:67}. To test if our catalog preferentially selects disk dwarfs ($\zeta \ge 0.825$) or subdwarfs ($\zeta < 0.825$), we again investigated the fraction of returned W11 stars after passing them through our color selection criteria, and results are shown in Table~\ref{tbl:matches}. 
	
	Our selection criteria are well-suited for retrieving both disk dwarf and subdwarf populations, with selection of stars in high extinction environments being slightly less reliable, but all above 92\%. Our ability to select the vast majority of spectroscopic low-mass stars based on colors alone demonstrate that these color criteria are suitable for selecting photometric low-mass stars, even in regions of moderate extinction.
	
\begin{deluxetable}{c c c c}
\tabletypesize{\footnotesize}
\tablecolumns{4}
\tablecaption{Color Selection Criteria Returns for W11\label{tbl:matches}}
\tablehead{
\colhead{Color Criteria} & \colhead{Total} & \colhead{Disk Dwarfs} & \colhead{Subdwarfs}
}
\startdata
\multicolumn{4}{c}{$A_V \leq 0.5$}\\ 
\hline
1\tablenotemark{a}		& 98.8\%	& 98.8\%	& 99.2\%\\
2\tablenotemark{b} 		& 99.2\%	& 99.2\%	& 98.7\%\\
3\tablenotemark{c}		& 99.6\%	& 99.6\%	& 95.8\%\\
1+2 					& 98.3\%	& 98.3\%	& 98.3\%\\
2+3			 		& 99.0\%	& 99.0\%	& 95.0\%\\
1+3 					& 98.5\%	& 98.6\%	& 95.8\%\\
1+2+3				& 98.1\%	& 98.1\%	& 95.0\%\\
\hline\\
\multicolumn{4}{c}{$A_V > 0.5$}\\
\hline
1\tablenotemark{a}		& 98.4\%	& 98.4\%	& 98.1\%\\
2\tablenotemark{b} 		& 97.2\%	& 97.2\%	& 96.3\%\\
3\tablenotemark{c}		& 97.0\%	& 97.2\%	& 94.4\%\\
1+2 					& 96.2\%	& 96.2\%	& 96.3\%\\
2+3			 		& 95.9\%	& 96.2\%	& 92.6\%\\
1+3 					& 96.1\%	& 96.2\%	& 94.4\%\\
1+2+3				& 94.9\%	& 95.2\%	& 92.6\%
\enddata
\tablenotetext{a}{$r-i$ vs. $i-z$ criteria.}
\tablenotetext{b}{$r-z$ vs. $z-J$ criteria.}
\tablenotetext{c}{$r-z$ vs. $z-W1$ criteria.}
\end{deluxetable}

\subsection{Initial Stellar Sample}\label{selection}

\begin{figure}
\centering
 \includegraphics{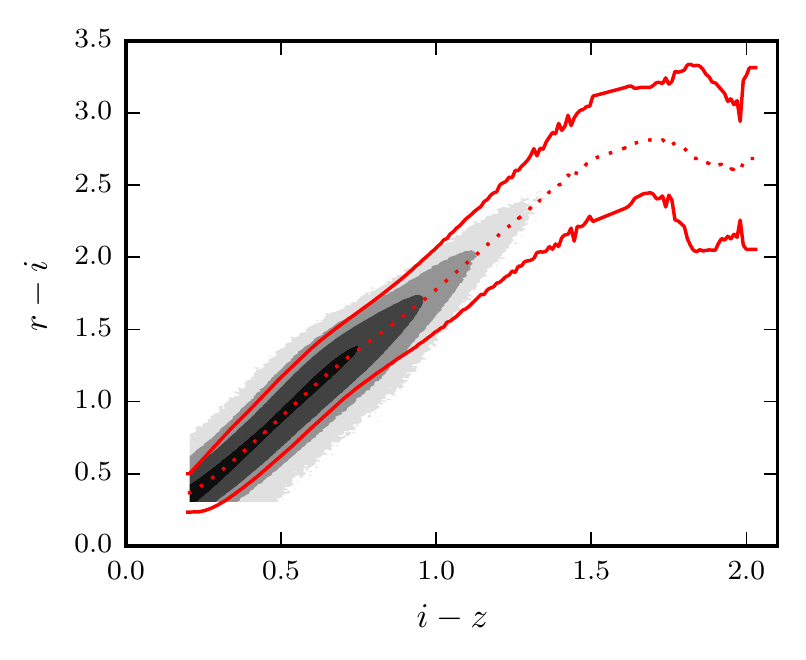}
 \includegraphics{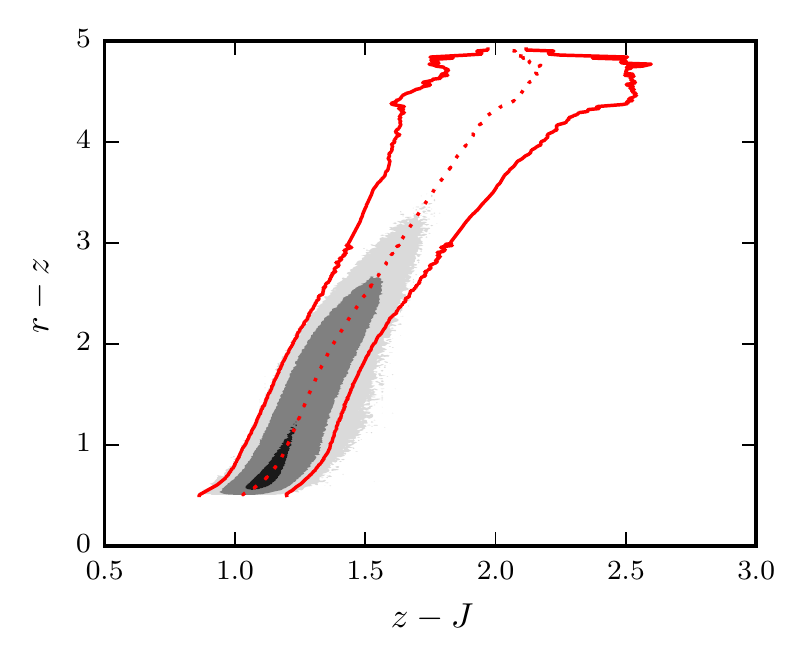}
 \includegraphics{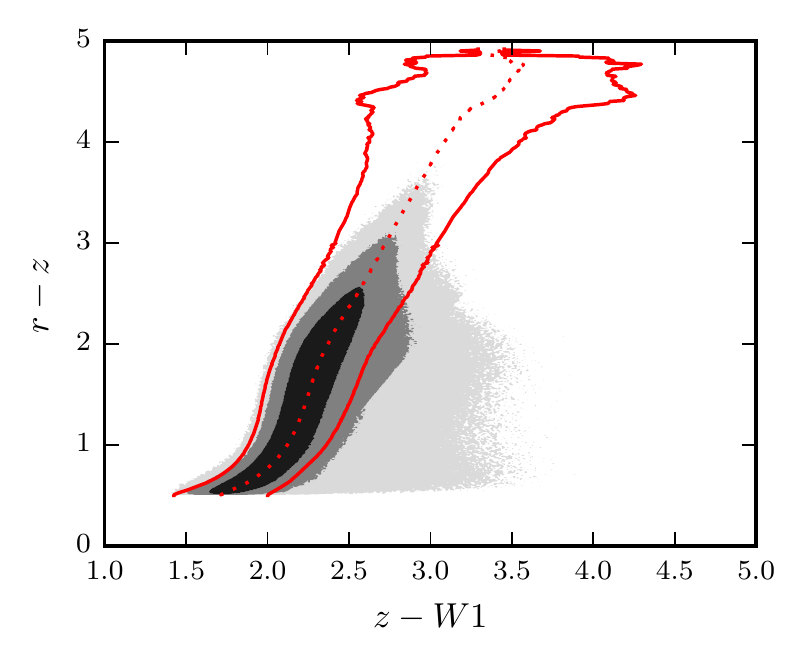}
\caption{Density plots in \textit{WISE}, SDSS, and 2MASS colors, each bin is (0.01 mag)$^2$. The first filled contour is drawn at 10 sources per bin, and each subsequent filled contour increases by a factor of 10. Mean colors and 3$\sigma$ errors used to select our initial stellar sample are also shown (dotted line and solid lines, respectively). Our method well approximates the stellar locus. The points used to draw the limits (dashed and dotted lines) are shown in Appendix~\ref{polys}.
\label{fig:ColorCuts}}
\end{figure}

\begin{figure}
\centering
 \includegraphics{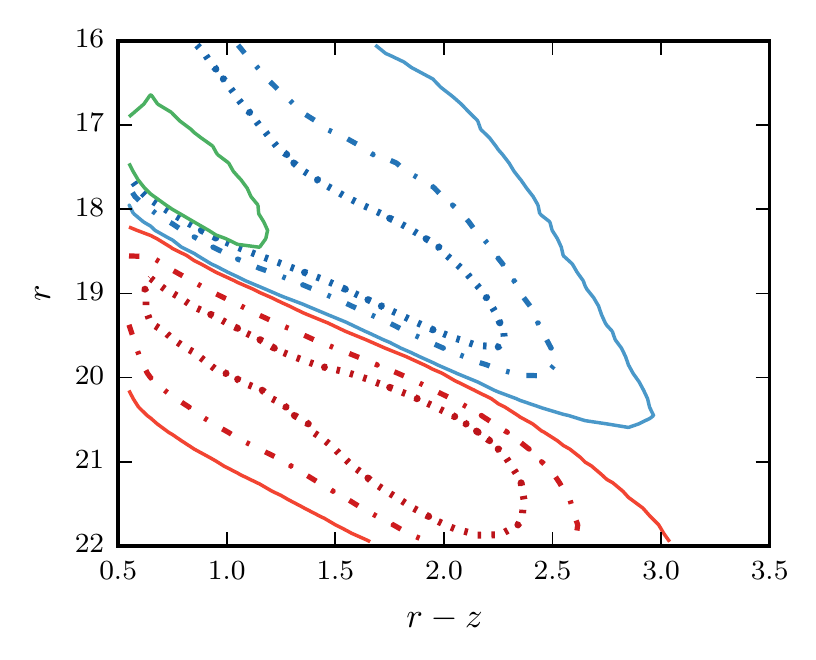}
\caption{Hess diagram ($r$ vs. $r-z$) with density contours, each bin is (0.1 mag)$^2$. \textit{WISE}+SDSS+2MASS are drawn in blue, \textit{WISE}+SDSS are drawn in red, and SDSS+2MASS are drawn in green. Contours are drawn at 1000, 5000, and 10000 stars per bin (solid, dash-dotted, and dashed lines, respectively). Each subsample is biased towards a specific brightness and color range.
\label{fig:hess}}
\end{figure}

	Applying the color cuts from Figure~\ref{fig:ColorCuts} to our initial stellar sample, we were left with 24,571,934 stars. The color and magnitude range over which each of our matched samples is found is shown in Figure~\ref{fig:hess}. \textit{WISE}+SDSS sources are typically fainter and redder, biasing these matches to later-type stars. SDSS+2MASS matches are bluer, biasing these matches to the earliest type stars in our sample.

	Using the above sources as our input catalog, we computed proper motions following Section~\ref{pmalg}. The surveys that went into each proper motion measurement (e.g., \textit{WISE}+SDSS+2MASS or \textit{WISE}+SDSS) are indicated in our catalog by a three bit flag (\textsc{dbit}), where each bit represents if a survey was used in the proper motion measurement. Descriptions of \textsc{dbit} and the number of initial sources with each fit are shown in Table~\ref{tbl:dbits}. 
	
\begin{deluxetable}{c l r}
\tabletypesize{\scriptsize}
\tablecolumns{3}
\tablewidth{0pt}
\tablecaption{Proper Motion Detection Flags\label{tbl:dbits}}
\tablehead{
\colhead{\textsc{DBIT}} & \colhead{Description} & \colhead{Number}
}
\startdata
011 & SDSS and \textit{WISE} surveys were used & $initial$: 11,911,109\\
 & to calculate the proper motions. & $final$: 1,801,369\\[2mm]
110 & 2MASS and SDSS surveys were used & $initial$: 1,052,228\\
 & to calculate the proper motions. & $final$: 69,199\\[2mm]
111 & All three surveys were available and & $initial$: 11,608,597\\
 & used in the linear fit to calculate & $final$: 1,733,104\\
 & proper motions. &\\[2mm]
 000 & SDSS+USNO-B proper motions & $final$: 6,620,838\\
 & are available. & $unique$: 5,216,855
\enddata
\tablecomments{We show both the initial number of sources that met our first selection criteria (see Section~\ref{data}), and the final number of stars that went into our catalog.}
\end{deluxetable}

\subsection{Reducing Contamination}\label{contam}

	There are two main issues that can contribute to spurious motion estimates for our stars: 1) nearby sources that can offset the center of the measured PSF if deblending methods fail; and 2) short time baselines that can contain motion caused by parallactic effects in addition to tangential motion. We will closely examine both of these issues in the following subsections.

\subsubsection{Nearby Neighbors}\label{neighbors}

\begin{figure}
\centering
 \includegraphics{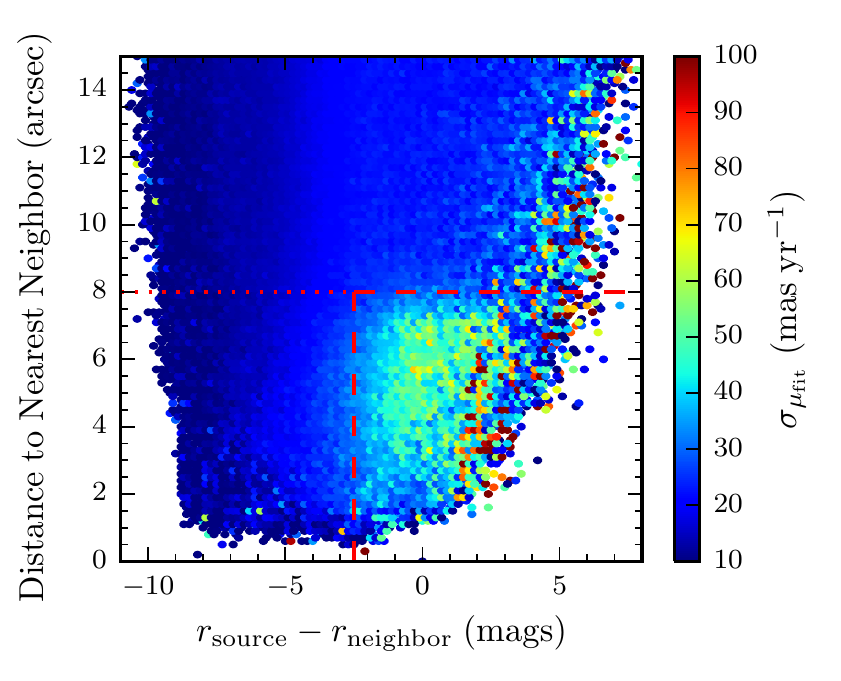}
\caption{Total proper motion fitting errors for the \textit{WISE}+SDSS+2MASS proper motions as a function of distance to nearest SDSS primary object and magnitude difference between our source and its closest neighbor. Only stars with one neighbor within the search radius are shown, each bin is 0.2 mags $\times$ 0.2\arcsec. Stars with fitting errors $> 40$ \masyear\ are suspect due to blending. Dashed and dotted lines correspond to criteria, below which we recomputed proper motions or removed sources (see text for further details). For \textit{WISE}+SDSS+2MASS stars, we recomputed proper motions for stars within the dashed lines. For \textit{WISE}+SDSS sources, we removed stars that had a neighboring object within $\leq 8$\arcsec\ (dotted line).
\label{fig:ndc}}
\end{figure}

	Close neighboring sources are expected to be a contaminant, especially for \textit{WISE} stars due to the large beam size (FWHM$_{W1}$=6.1\arcsec). \textit{WISE} active deblending allows, at most, two components to the PSF fit. The robustness of this deblending is likely dependent on the flux difference between the two blended sources. To investigate how neighboring objects affect proper motion measurements, we used SDSS CasJobs to select all neighboring primary objects within 15\arcsec\ of our sources' SDSS positions. Next, we selected only sources that had one neighboring object within the 15\arcsec\ search radius. We expect sources whose photocenters have been moved significantly in \textit{WISE} will have larger fitting errors for sources detected in all three surveys. Figure~\ref{fig:ndc} shows fitting errors as a function of distance to a neighboring object and $r$-band magnitude difference between our source and the neighboring object. Fitting errors: (1) are significantly larger for brighter neighboring objects, and objects at a distance $\lesssim 8\arcsec$; and (2) decrease for extremely close (and bright) neighboring objects ($\lesssim 2\arcsec$) due to a reduction in the offset of the measured photocenter from its true position, making the measured positional deviation small.
	
	Figure~\ref{fig:ndc} illustrates that sources with neighboring, bright objects have problematic motion estimates that can be seen in the fitting errors. The following questions must be answered to determine which proper motion measurements are reliable: 1) at what distance does a neighboring object affect our measurement; 2) at what magnitude difference does a neighboring object affect our measurement; and 3) what is the fitting error threshold above which a measurement is considered affected (by either neighbors or parallactic effects)? We will explore each of these three questions below.
	
	We expect the error distributions for sources with a neighboring object at a distance $\lesssim 8$\arcsec\ and $> 8$\arcsec\ to be different due to the inclusion of more large errors for the distribution with a neighboring object $< 8$\arcsec. We require a statistical argument in choosing at what distance a neighboring object affects our measurement. To test for similarity (or difference) in the error distributions, we chose to use the Anderson-Darling test since it is more sensitive to the tails of the distribution, where we expect the larger fitting errors to reside. To determine at which distance a neighboring object affects our proper motion measurement, we selected all sources with $r_{\rm source} - r_{\rm neighbor} > -2$. The requirement for sources with $r_{\rm source} - r_{\rm neighbor} > -2$ is relatively arbitrary, we are selecting sources where we know fitting errors begin to increase, later we make a more rigorous estimate for $r_{\rm source} - r_{\rm neighbor}$. Next, we binned fitting errors in steps of 0.5\arcsec\ between 4\arcsec\ to 12\arcsec, performing an Anderson-Darling test between adjacent 0.5\arcsec\ bins, searching for neighboring distributions with the largest dissimilarity (as traced through a minimum in the p-value). We found a minimum p-value ($8 \times 10^{-6}$) between the bins from 7.5\arcsec\ to 8\arcsec\ and 8\arcsec\ to 8.5\arcsec, making 8\arcsec\ our cutoff value.
	
	Next, to determine at which magnitude difference ($r_{\rm source} - r_{\rm neighbor}$) a neighboring object affects our proper motion measurement, we selected all sources with a neighboring object $\leq 8\arcsec$. Again, we binned fitting errors in steps of 0.5 mag between $-5$ to 3 mags, and performed Anderson-Darling tests between adjacent distributions. We found a minimum p-value ($4 \times 10^{-7}$) between the bins from $r_{\rm source} - r_{\rm neighbor}= -3$ to $-2.5$ and $-2.5$ to $-2$, making $-2.5$ our cutoff value. Additionally, we visually inspected a number of stars with proper motion errors between $\sigma_\mu = 30$--50 \masyear\ and found that sources with $\sigma_\mu > 40$ \masyear\ tended to be spurious measurements. We chose to recompute proper motions for these stars using just two of the three surveys. Details regarding which surveys were chosen (\textit{WISE}+SDSS or SDSS+2MASS) are described below.
	
	Our method for recomputing proper motions is as follows:
\begin{enumerate}
\item We completely removed stars with a neighboring source found within 8\arcsec\ and with $r_{\rm source} - r_{\rm neighbor} > -2.5$. 
\item For stars that do not meet the above criteria, but which have $\sigma_\mu > 40$ \masyear\ in either the $\alpha$ or $\delta$ component, we recomputed proper motions by first requiring a minimum signal-to-noise (S/N) in each survey. For \textit{WISE}, we required S/N$_{W1} \geq 3$, and for 2MASS, we required S/N$_{J,H,\; {\rm or}\; K_s} \geq 5$. If only one survey met our minimum S/N threshold, we use that survey. If both surveys met our S/N threshold, we used the survey with the highest S/N. To use the \textit{WISE} baseline, we further required a minimum distance to the nearest neighbor be $\geq 8$\arcsec. 
	\begin{enumerate}
	\item If all the previous criteria of (2) were met, and the highest S/N between both surveys was equal, we used the two surveys with the longest time baseline (either \textit{WISE}+SDSS or SDSS+2MASS). 
	\end{enumerate}
\item Lastly, we required a time baseline $> 1$ year between observations. Time baselines shorter than this are susceptible to parallax effects for nearby stars ($\sim$20 \masyear\ for a star at 100 pc with a baseline of 6 months).
\end{enumerate}
		
	As a final note, the buildup of large errors at a neighboring distance of $\sim$6\arcsec\ may also be due to our search radius (6\arcsec) picking up neighboring objects as the primary if our source becomes too faint at longer wavelengths. Removing these sources ensures both possibilities for contamination are removed from our catalog. Since we are only interested in stars exhibiting \textit{bona-fide} tangential motions, we kept only stars with total proper motions greater than twice their uncertainty ($\mu_{\rm tot} > 2\sigma_{\mu_{\rm tot}}$).

\section{Reliability of Proper Motions}\label{reliability}

	To assess the reliability of our catalog, we compared our proper motions to the LSPM catalog, the SDSS+USNO-B catalog \citep[][hereafer M04]{munn:2004:3034, munn:2008:895}, and the recent deep survey completed within a 1098 deg$^2$ SDSS footprint \citep[][hereafter M14]{munn:2014:132}. The reliability of LSPM should be close to 100\% as all these sources have been verified by eye. LSPM stars are selected for larger proper motions ($>150$ \masyear), and the catalog is biased towards brighter stars due to the use of Schmidt plates for the earliest baselines. M04 has time baselines of $\sim$50 years, giving it a high precision, but M04 is not as deep as our catalog, also due to the use of Schmidt plates. M04 has also been matched to a high number of SDSS sources since SDSS was used as the most recent baseline in computing proper motions. Lastly, M14 allows us to test the fidelity of our faintest sources, which LSPM and M04 do not probe. Together, these catalogs allow us to assess the reliability of our sources for the stars with both small and large proper motions, and across all magnitudes.

\subsection{Comparison to LSPM}\label{LSPM}

\begin{figure*}
\centering
 \includegraphics[width=.8\textwidth]{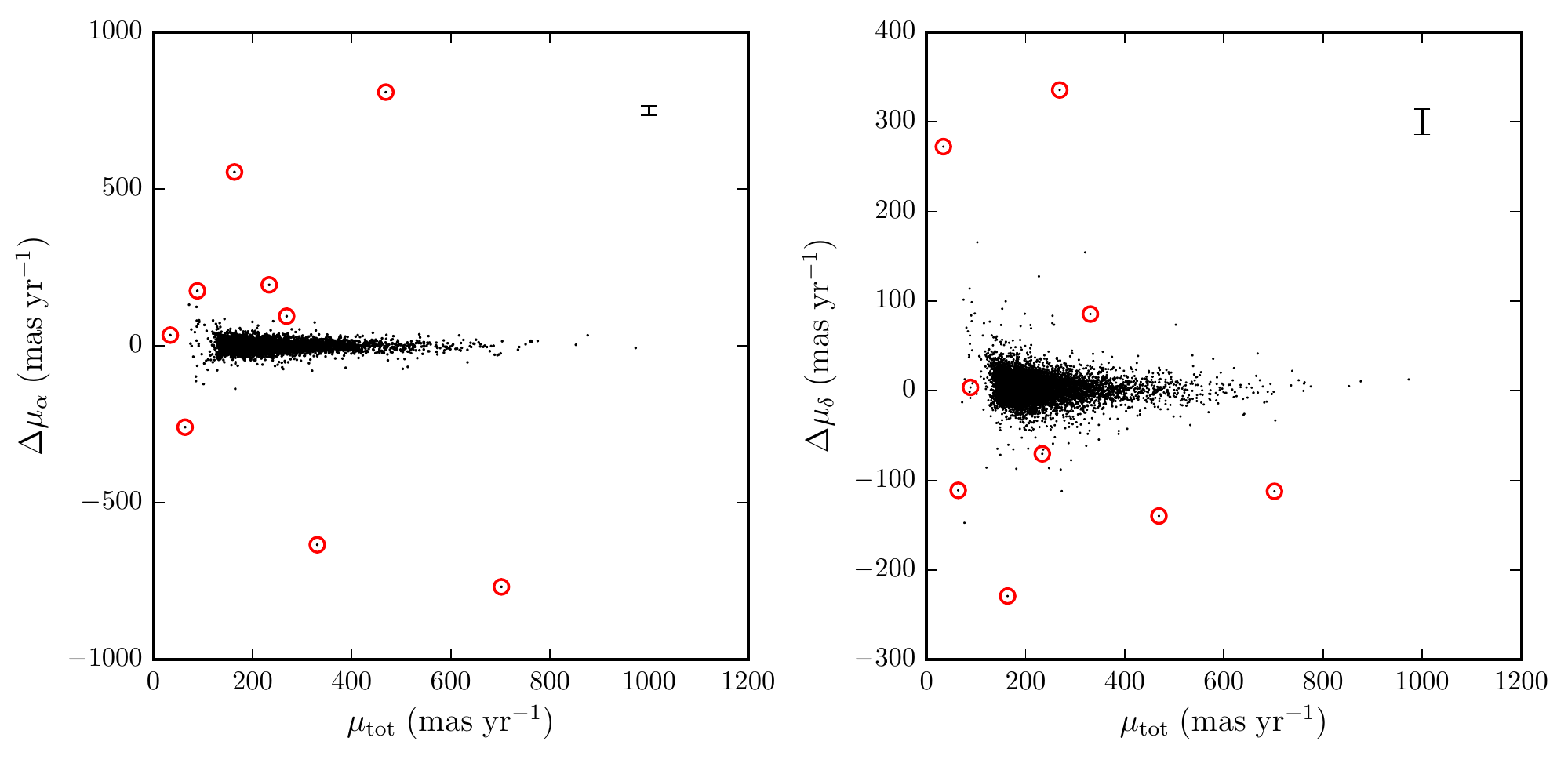}
 \includegraphics[width=.8\textwidth]{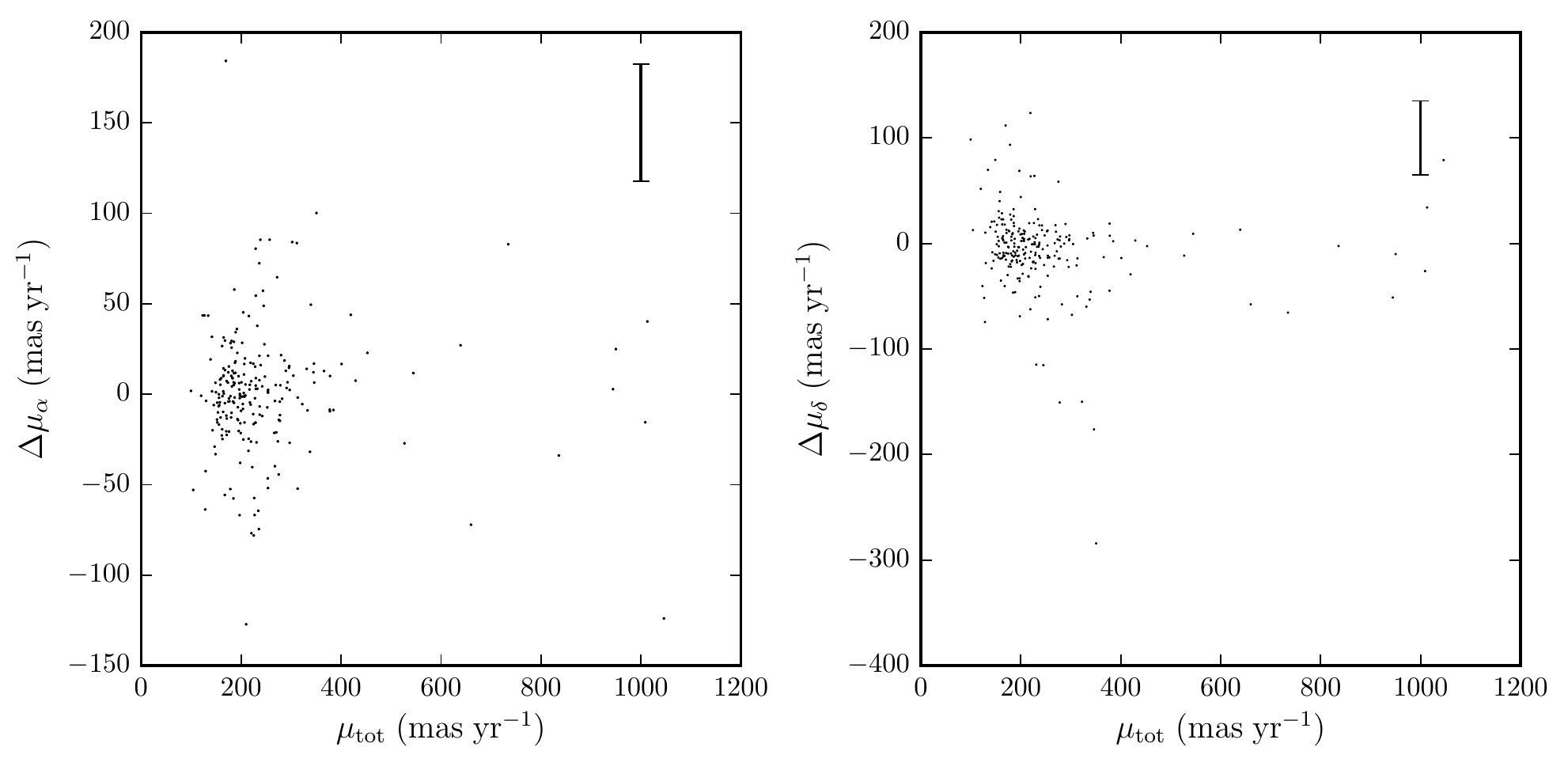}
 \includegraphics[width=.8\textwidth]{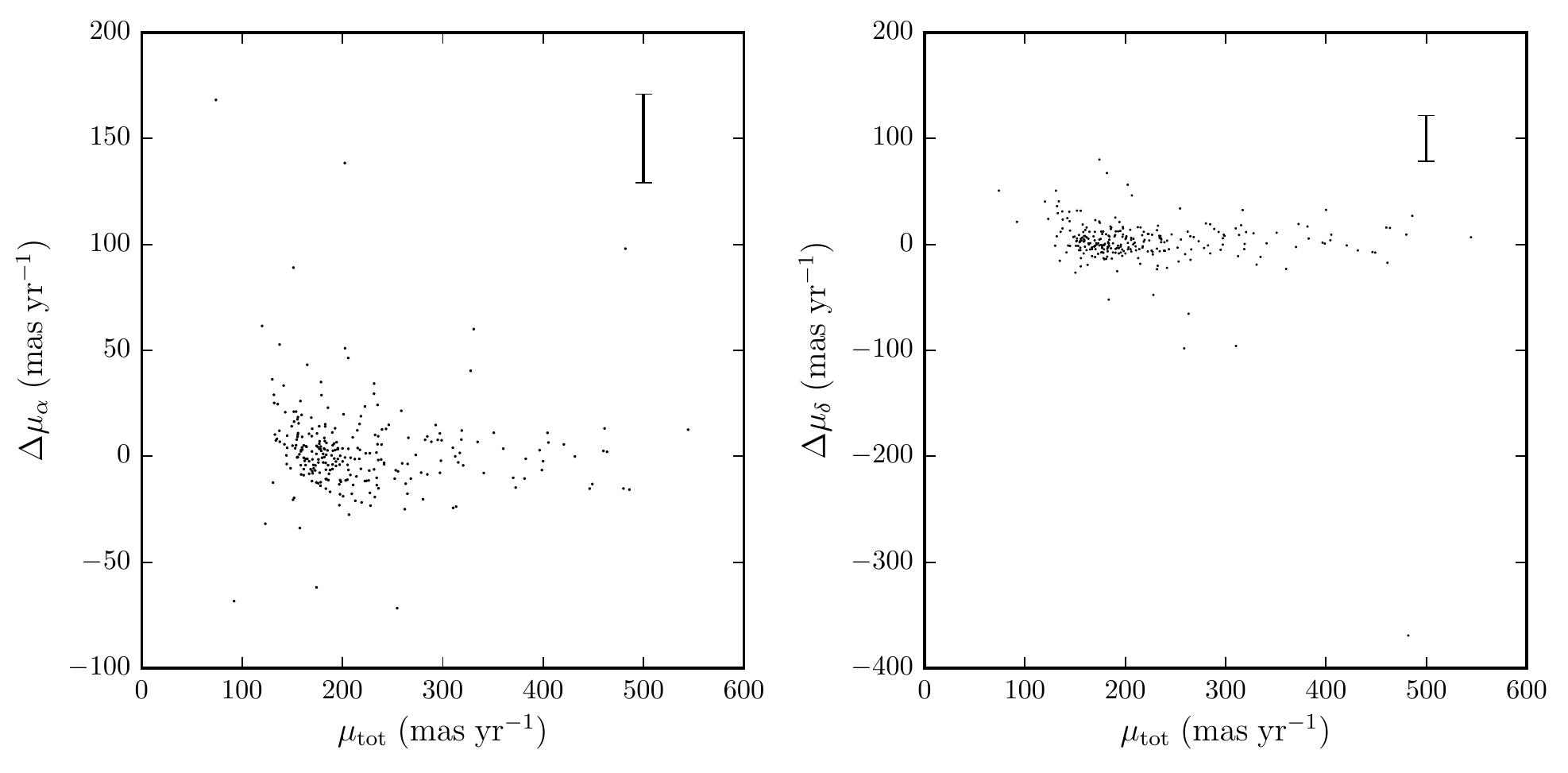}
\caption{Residual proper motion ($\mu_\textrm{this study} - \mu_\textrm{LSPM}$) for both components ($\mu_\alpha$ and $\mu_\delta$) as a function of our total proper motion. Typical errors are shown in the top right corners. \textit{Top}: \textit{WISE}+SDSS+2MASS stars compared against LSPM. Overall, our stars show good agreement with LSPM, however, a small number of outliers were identified and investigated (red circles, see text for details). \textit{Middle}: \textit{WISE}+SDSS stars compared to LSPM. Again, there is good agreement between our stars and LSPM, with the outliers typically due to bad astrometry in one of our surveys. \textit{Bottom}: SDSS+2MASS stars compared to LSPM. Similar to our \textit{WISE}+SDSS stars there is good agreement with LSPM, with the outliers once again due to bad astrometry in one of our surveys.
\label{fig:LSPM}}
\end{figure*}

	The LSPM catalog contains 61,977 stars in the northern hemisphere with proper motions $> 150$ \masyear. The precision of LSPM is $\sim$8 \masyear. LSPM was not specifically designed to target low-mass stars, however, due to its selection of high proper motion stars, it primarily consists of nearby dwarf stars. We matched our catalog to LSPM stars using their 2MASS designations, which produced 12,930 matches. We investigated the agreement between our catalog and LSPM in all three subsamples (e.g., \textit{WISE}+SDSS+2MASS) for each proper motion component, as shown in Figure~\ref{fig:LSPM}. A small number of large outliers were identified between our \textit{WISE}+SDSS+2MASS and LSPM matches (Figure~\ref{fig:LSPM}, red circles). We looked through archived images (DSS, SDSS, 2MASS, and \textit{WISE}), since these stars should all have apparent proper motions over the $\gtrsim 20$ year baseline in archived images. The majority of these stars showed proper motions more consistent with our measurements in the archived images. The cause of the spurious measurement in LSPM is unclear, presumably these are bad residual images in the SUPERBLINK pipeline that eluded inspection. Many of these stars also had proper motions consistent with our measurement in another catalog (e.g., USNO-B, M04, NLTT, PPMXL, or URAT1).

	We compared the 2$\sigma$ agreement in both proper motion components to test the absolute agreement between our catalog and LSPM. For our \textit{WISE}+SDSS+2MASS matches, we found 98\% agreement between our catalog and LSPM. For our \textit{WISE}+SDSS and SDSS+2MASS matches this agreement was 97\% and 96\%, respectively. All agreements increase to 99\% at 3$\sigma$. All our \textit{WISE}+SDSS matches to LSPM were recomputed values (see Section~\ref{neighbors}), since these stars are all bright enough to have an entry in 2MASS. This comparison shows our catalog is reliable for the fastest moving stars.

\subsection{Comparison to M04}\label{M04}
	
	Due to the sensitivity of M04, spurious proper motion estimates are expected for fainter stars. To compare our catalog to M04 we needed to determine which sources were reliable. We matched our catalog to the \citet{munn:2004:3034} catalog (contained within SDSS CasJobs as the \textsc{ProperMotions} table) by SDSS \textsc{objid}. We chose to use only sources with ``good" proper motions, adopting the constraints from \citet[][]{kilic:2006:582}:
\begin{enumerate}
\item \textsc{dist22} $>7$, the nearest neighboring objects with $g<22$ is more than 7\arcsec\ away.
\item \textsc{match} $=1$, there is a one-to-one match with the USNO-B object and the SDSS object.
\item \textsc{nFit} $=6$, the object was detected in all five USNO-B plates, as well as detected in SDSS
\item \textsc{sigRA} $<525$, and \textsc{sigDEC} $<525$, the RMS residuals for the proper motion fits were less than 525 mas in both components.
\end{enumerate}
These criteria yielded 919,867 matches between our SDSS sources and M04 sources.

\begin{figure}
\centering
 \includegraphics{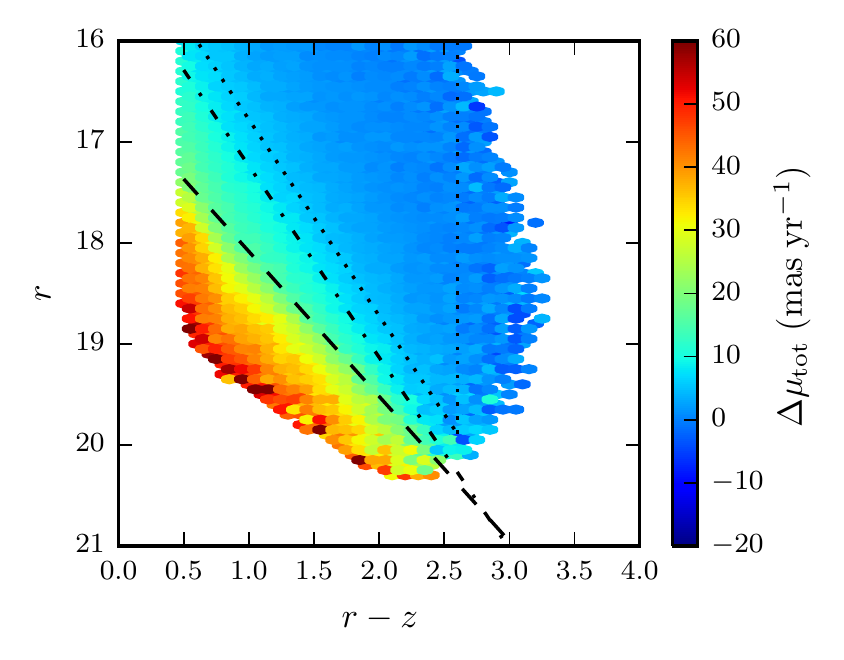}
 \includegraphics{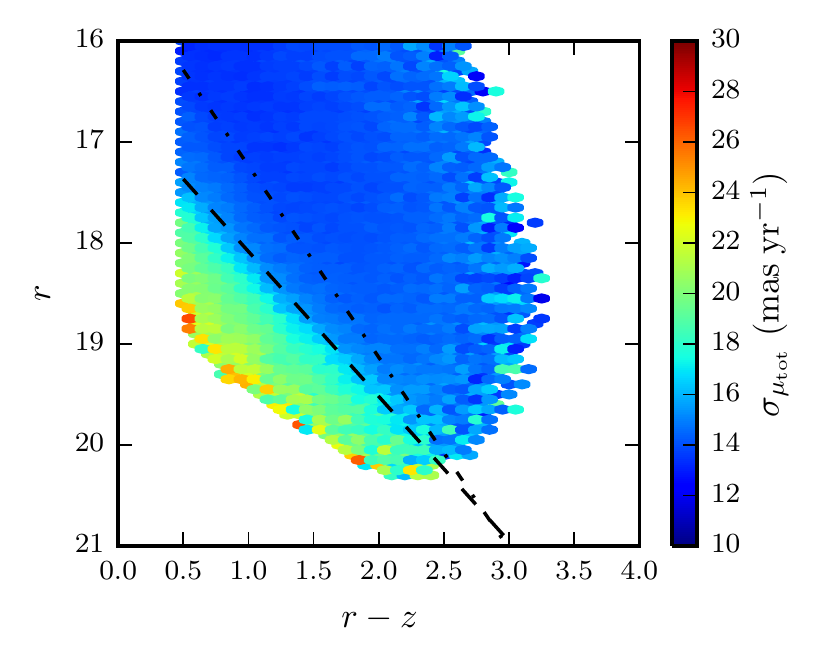}
\caption{\textit{Top}: Residual total proper motion ($\Delta \mu_{\rm tot} =$ $\mu_{{\rm tot}\; (WISE+{\rm SDSS+2MASS})} - \mu_{\rm tot\; (M04)}$ as a function of source $r$-band magnitude and $r-z$ color, each bin is (0.1 mags)$^2$. The performance of our catalog is tightly correlated with source color and magnitude. The dashed line represents the limit above which residuals become $> 20$ \masyear, and the dash-dotted line represents the limit above which residuals become $> 10$ \masyear. More stringent criteria used to investigate the reliability of the M04 catalog (Section~\ref{usno}) are also shown (dotted lines).
\textit{Bottom}: Total proper motion error ($\sigma_{\mu_{\rm tot}}$) as a function of source $r$-band magnitude and $r-z$ color, each bin is (0.1 mags)$^2$. The dashed and dash-dotted lines are the same as the top. Our measurement uncertainty becomes large at the same limit that residuals begin to grow larger than 20 \masyear. The performance of our catalog is tightly correlated with source color and magnitude, due to our inability to probe small proper motions expected for more distant (fainter) sources. Although we lose precision at fainter magnitudes, our proper motion measurements are still consistent to within the errors.
\label{fig:colormag}}
\end{figure}

	Since M04 is biased towards brighter objects, we chose to compare it to sources that have proper motions measured using all three surveys (\textit{WISE}+SDSS+2MASS). There were 842,776 matched stars with \textit{WISE}+SDSS+2MASS measured proper motions, making up the bulk of the matches between our catalog and M04. We investigated the residuals between our computed proper motions and those from M04 as a function of color and magnitude (Figure~\ref{fig:colormag}). There is a strong increase in the magnitude of the residuals for fainter stars and bluer stars. The correlation between apparent magnitude and $r-z$ color is due to bluer stars peaking in the optical (and NIR), while getting fainter at \textit{WISE} bands. The combined filter set of \textit{WISE}+SDSS+2MASS is more sensitive to stars that peak towards the red end of the NIR, out to $\sim$ 4 \um. We expect M04 to be more sensitive to bluer stars, however, we can examine the proper motion distributions for these faint sources to determine whether ours, or the M04 values are more consistent with expectations.
	
	In general, we expect fainters stars to be more distant, and therefore exhibit smaller proper motions. Our proper motions for fainter sources tend to be higher than those from M04, as can be see in Figure~\ref{fig:colormag} (positive residuals). We performed a linear fit to the residuals in color-magnitude space, fitting to where the residuals became higher than 20 \masyear (dashed line, Figure~\ref{fig:colormag}), and where the residuals became higher than 10 \masyear (dash-dotted line, Figure~\ref{fig:colormag}). Our results are shown in Figure~\ref{fig:colormag}, with the proper motions separated by,
\begin{equation}\label{cut1}
r \leq 16.65 + 1.435(r-z) \hspace{10pt} \text{for } \Delta \mu_\mathrm{tot} < 20 \text{ mas yr}^{-1},
\end{equation}
and
\begin{equation}\label{cut2}
r \leq 15.34 + 1.896(r-z) \hspace{10pt} \text{for } \Delta \mu_\mathrm{tot} < 10 \text{ mas yr}^{-1}.
\end{equation}

	As stated above, due to the difference in wavelengths between each of the surveys used in calibrating our catalog, we expect fainter, redder sources to be more easily detectable than fainter, bluer sources. This is exhibited in Figure~\ref{fig:colormag}, where our proper motion errors also begin to grow large at approximately the same limit that our residuals began to increase. Therefore, both our proper motions and our proper motion errors increase at the same limit, potentially keeping our measurements to within errors but with less precision. To quantify the reliability of our catalog, we compared the 2$\sigma$ and 3$\sigma$ agreement for both proper motion components for matched stars, both fainter and brighter than equation~(\ref{cut1}), and as a function of color, the results of which are shown in Table~\ref{tbl:PMagreement}.
	
	In general, there is good agreement between our measurements and the M04 catalog. This agreement is less certain for stars fainter than equation~(\ref{cut1}), but increases for redder colors fainter than this limit. Since the performance of the USNO-B survey is also expected to deteriorate at fainter magnitudes \citep[$V\gtrsim19$;][]{lepine:2005:1483, dong:2011:116}, we can expect sources in disagreement to be a mixture of spurious measurements in our catalog and in USNO-B. This is left as a cautionary note, as many of the stars fainter than equation~(\ref{cut1}) will have real proper motions. We do not remove these stars, as many of them will be true measurements, we instead augment them with a second measurement from M04 (see Section~\ref{usno}).

\begin{deluxetable*}{c c c c c}
\tabletypesize{\footnotesize}
\tablecolumns{5}
\tablewidth{\textwidth}
\tablecaption{Agreement with other Proper Motion Catalogs\label{tbl:PMagreement}}
\tablehead{
\colhead{Comparison Catalog}	& \colhead{Subsample}	& \colhead{$r-z$}	& \colhead{2$\sigma$ Agreement}	& \colhead{3$\sigma$ Agreement}
}
\startdata
M04		& \textit{WISE}+SDSS+2MASS 	& All		& 86\%	& 95\%\\
M04		& SDSS+2MASS				& All		& 28\%	& 72\%\\
M04		& \textit{WISE}+SDSS			& All		& 15\%	& 50\%\\
M14		& \textit{WISE}+SDSS+2MASS		& All		& 82\%	& 94\%\\
M14		& SDSS+2MASS				& All		& 21\%	& 74\%\\
M14		& \textit{WISE}+SDSS			& All		& 25\%	& 63\%\\
M14		& \textit{WISE}+SDSS			& $>3$	& 82\%	& 93\%\\
\hline\\
\multicolumn{5}{c}{$r \leq 16.65 + 1.435(r-z)$}\vspace{2 pt}\\
\hline\\
M04		& \textit{WISE}+SDSS+2MASS 		& All		& 94\%	& 99\%\\
M04		& SDSS+2MASS				& All		& 87\%	& 97\%\\
M04		& \textit{WISE}+SDSS			& All		& 46\%	& 63\%\\
M14		& \textit{WISE}+SDSS+2MASS		& All		& 95\%	& 99\%\\
M14		& SDSS+2MASS				& All		& 69\%	& 95\%\\
M14		& \textit{WISE}+SDSS			& All		& 57\%	& 77\%\\
\hline\\
\multicolumn{5}{c}{$r > 16.65 + 1.435(r-z)$}\vspace{2 pt}\\
\hline\\
M04		& \textit{WISE}+SDSS+2MASS		& All		& 46\%	& 75\%\\
M04 		& \textit{WISE}+SDSS+2MASS		& $>1$	& 50\%	& 77\%\\
M04 		& \textit{WISE}+SDSS+2MASS		& $>1.5$	& 52\%	& 78\%\\
M04 		& \textit{WISE}+SDSS+2MASS		& $>2$	& 55\%	& 80\%\\
M04		& SDSS+2MASS				& All		& 13\%	& 66\%\\
M04		& \textit{WISE}+SDSS			& All		& 13\%	& 49\%\\
M04		& \textit{WISE}+SDSS			& $>2$	& 28\%	& 59\%\\
M14		& \textit{WISE}+SDSS+2MASS		& All		& 60\%	& 84\%\\
M14		& \textit{WISE}+SDSS+2MASS		& $>3$	& 93\%	& 98\%\\
M14		& \textit{WISE}+SDSS			& $>3$	& 80\%	& 92\%
\enddata
\end{deluxetable*}

	A similar comparison can be made between our \textit{WISE}+SDSS and SDSS+2MASS proper motions and M04. However, these samples are limited in that they are typically bluer (SDSS+2MASS) or fainter (\textit{WISE}+SDSS) than our \textit{WISE}+SDSS+2MASS stars. These comparisons are also shown in Table~\ref{tbl:PMagreement}. In general, agreement is again better for stars brighter than equation~(\ref{cut1}), and for redder stars. 
	
\begin{figure*}
\centering
 \includegraphics[width=.8\textwidth]{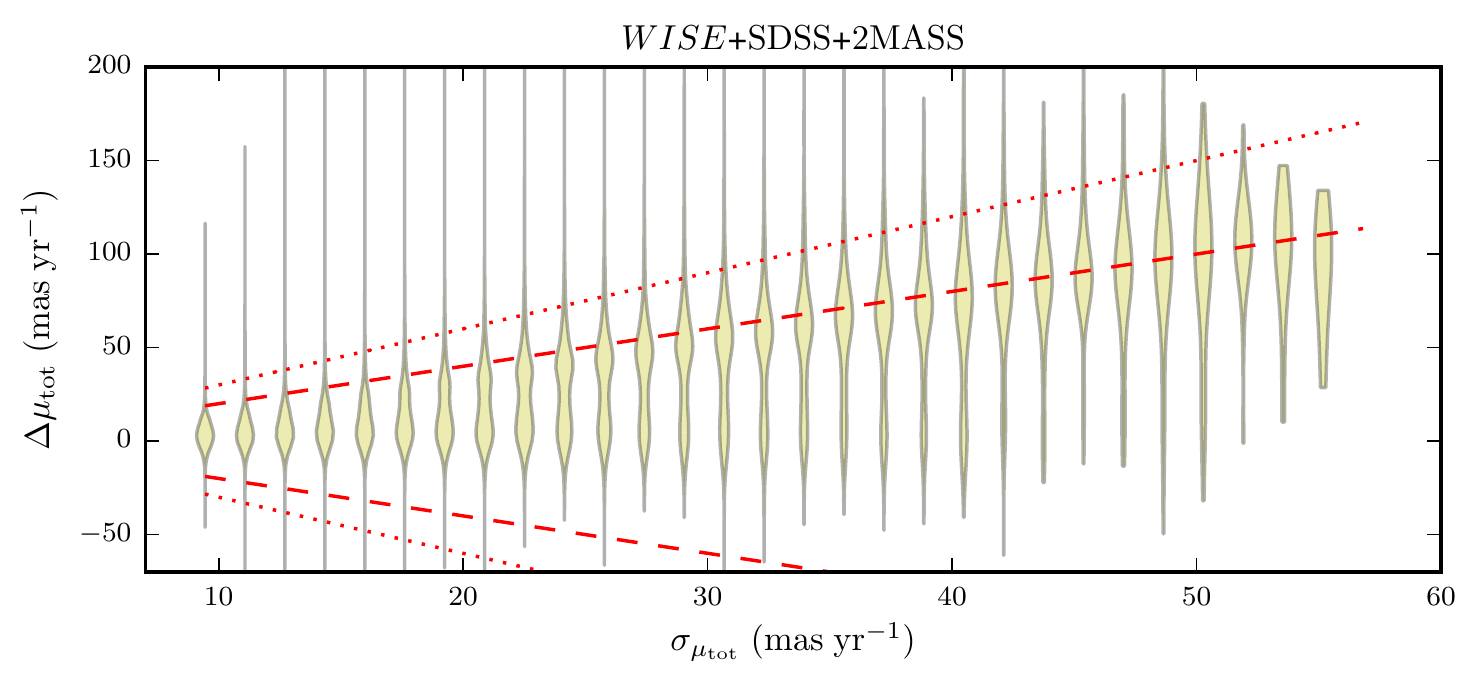}
 \includegraphics[width=.8\textwidth]{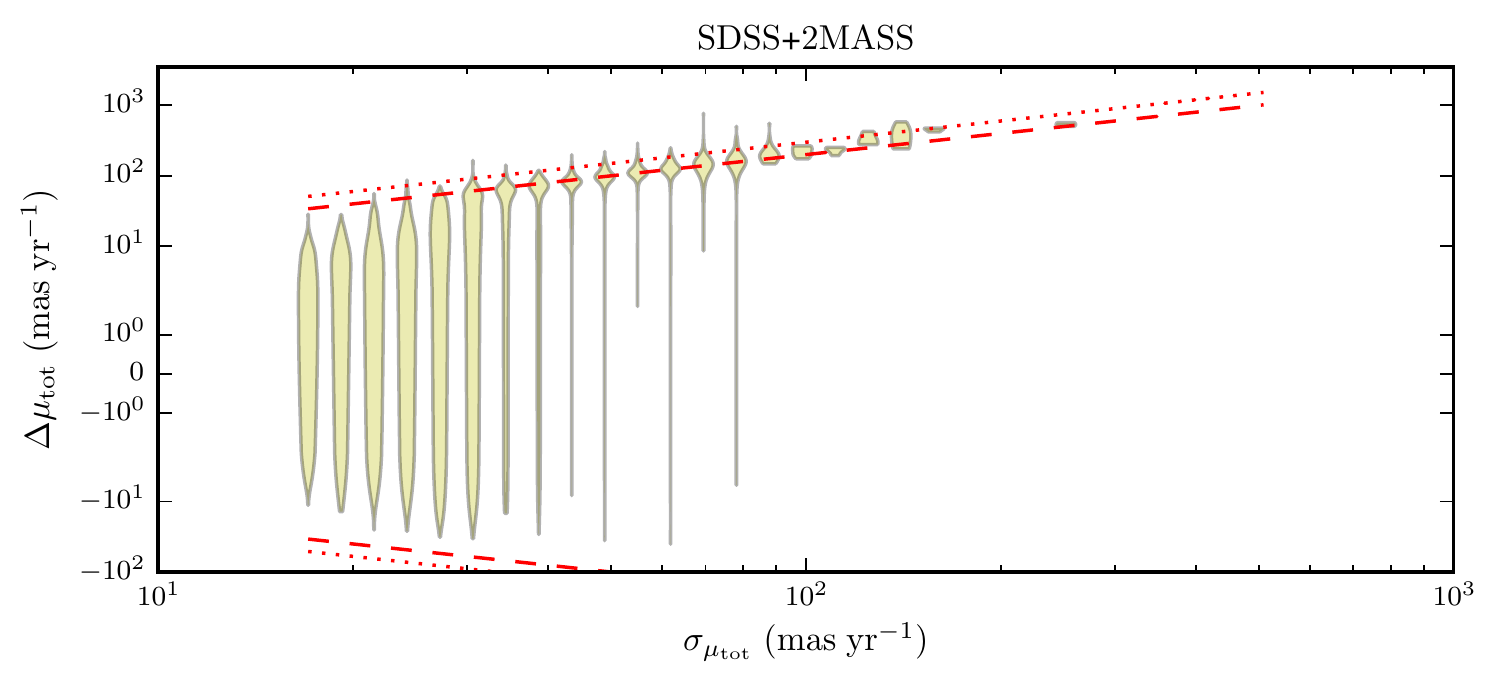}
 \includegraphics[width=.8\textwidth]{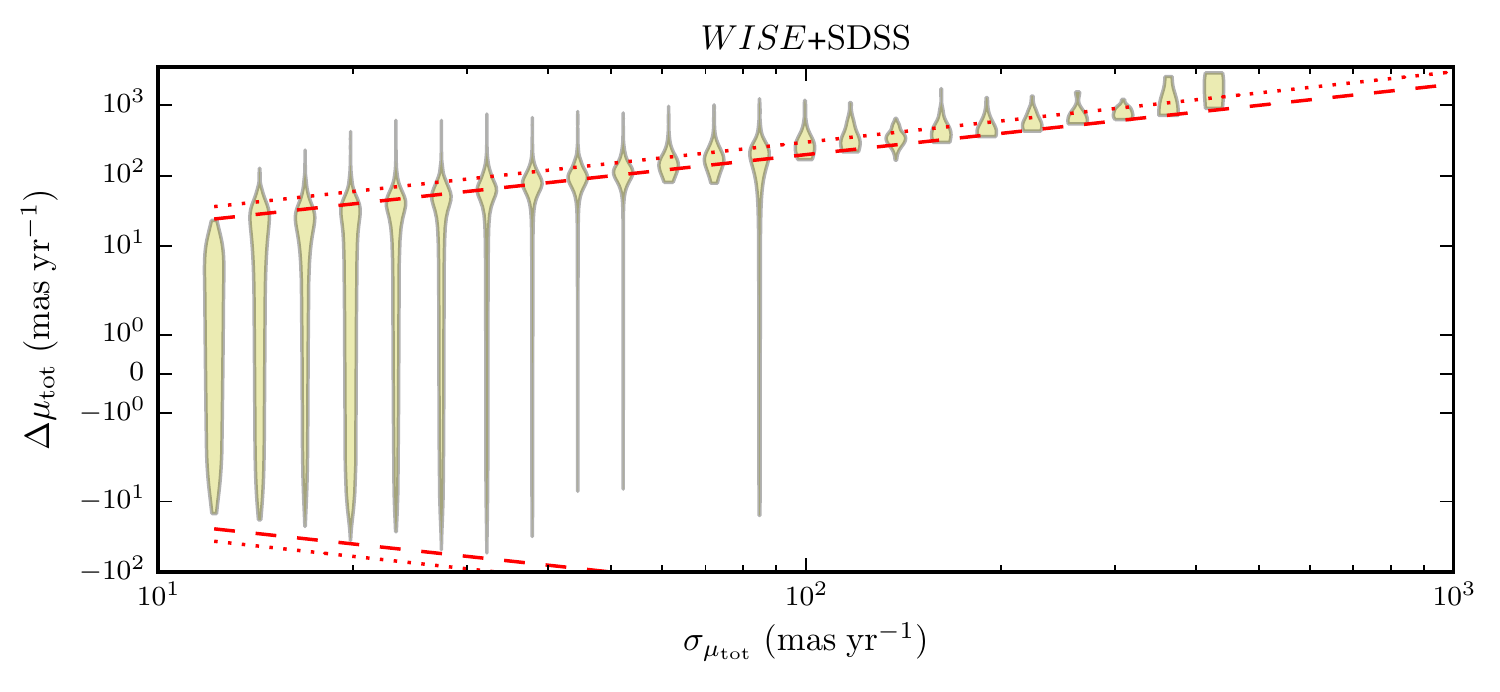}
\caption{Violin plots for residual total proper motion ($\Delta \mu_{\rm tot} =$ $\mu_{{\rm tot}\; (WISE+{\rm SDSS+2MASS})} - \mu_{\rm tot\; (M04)}$) as a function of total proper motion error. These plots show the relative distribution of stars for each error bin. The dashed and dotted line represent 2$\sigma$ and 3$\sigma$ agreement, respectively. \textit{Top}: The majority of \textit{WISE}+SDSS+2MASS stars are all in 3$\sigma$ agreement with M04, agreement is better for small errors. \textit{Middle}: The SDSS+2MASS stars tend to be in better agreement for smaller errors. From our error distribution (Section~\ref{movers}), the majority of our sources should have reliable (within 3$\sigma$) proper motions. \textit{Bottom}: The \textit{WISE}+SDSS stars also tend to be in better agreement for smaller errors. From our error distribution (Section~\ref{movers}), many sources tend to straddle the limit of 3$\sigma$ reliability. These stars also represent the faintest sources, and thus suffer from small number statistics when matched to M04.
\label{fig:M04reserr}}
\end{figure*}

	We further investigated how our agreement scaled with proper motion errors among all three subsamples, as is shown in Figure~\ref{fig:M04reserr}. Stars with larger proper motion errors (e.g., fainter stars and stars with smaller time baselines) tend to have to be closer to our 3$\sigma$ limit, or be in disagreement. The majority of stars across all three subsamples tend to be within 3$\sigma$ agreement, a proper motion error cut may be useful in selecting higher confidence proper motions. It is difficult to say if the disagreement between the faintest sources in our catalog (and M04) is a limitation of M04 or our catalog. To answer this question, we require a proper motion catalog as deep (or deeper) than our catalog.
	
\subsection{Comparison to M14}\label{M14}
	
	The M14 proper motion survey covers only a small footprint within SDSS, but provides proper motions down to $r \approx 22$. This catalog has time baselines $\gtrsim 5$ years with a precision of $\sim$10 \masyear. We applied all the M14 suggested quality cuts (summarized in Table 1 of M14). Although these cuts were designed to remove spurious proper motion measurements, M14 found that they were only able to remove $\sim$50\% of the ``bad" detections, while retaining $>$97.6\% of the ``good" detections (from comparison stars between M14 and LSPM). Therefore, it is possible that some of the proper motions retrieved are spurious measurements. The fidelity of the proper motions within this catalog have not been independently verified, therefore, the following analysis should be taken as both an investigation into the reliability of our catalog as well as M14. 
	
\begin{figure}
\centering
 \includegraphics{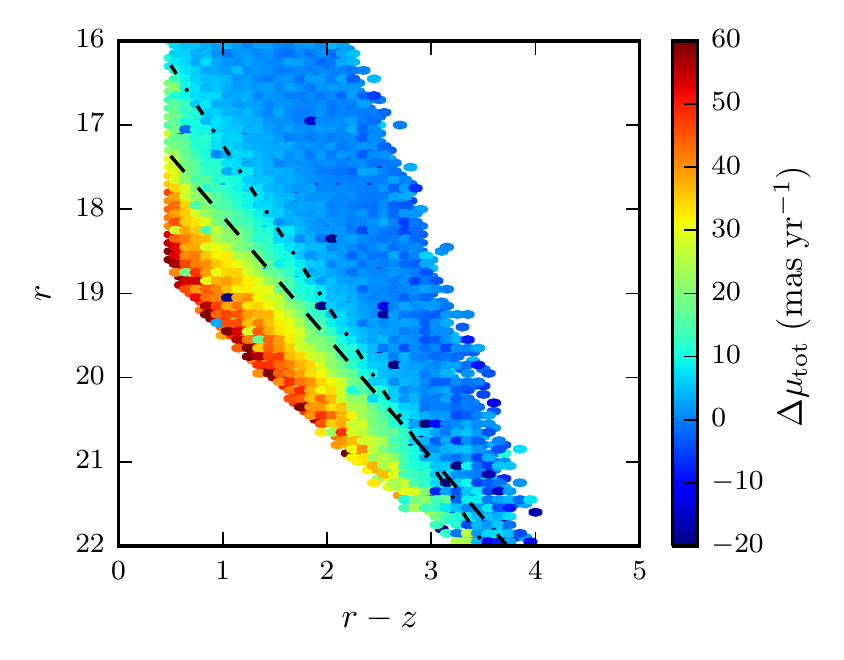}
\caption{Same as Figure~\ref{fig:colormag}, but residuals with M14. Equations~(\ref{cut1}) and (\ref{cut2}) approximately trace the same levels of reliability for M04 and M14, with reliability increasing for fainter, redder stars ($r-z > 3$).
\label{fig:colormag14}}
\end{figure}

	We matched our catalog to M14 by SDSS \textsc{objid}, giving us 510,799 matches. We performed an analysis similar to our comparison to M04 above. We began with a comparison of our \textit{WISE}+SDSS+2MASS stars, which made up 246,678 of our matches. Figure~\ref{fig:colormag14} shows the total proper motion residuals ($\mu_{{\rm tot}\; (WISE+{\rm SDSS+2MASS})} - \mu_{\rm tot\; (M14)}$) as a function of $r$-band magnitude and $r-z$ color. Equations~(\ref{cut1}) and (\ref{cut2}) from our M04 analysis represent similar reliability for M14, past the magnitude limit that M04 was able to probe. Reliability tends to increase for stars with $r-z > 3$.

\begin{figure*}
\centering
 \includegraphics[width=.8\textwidth]{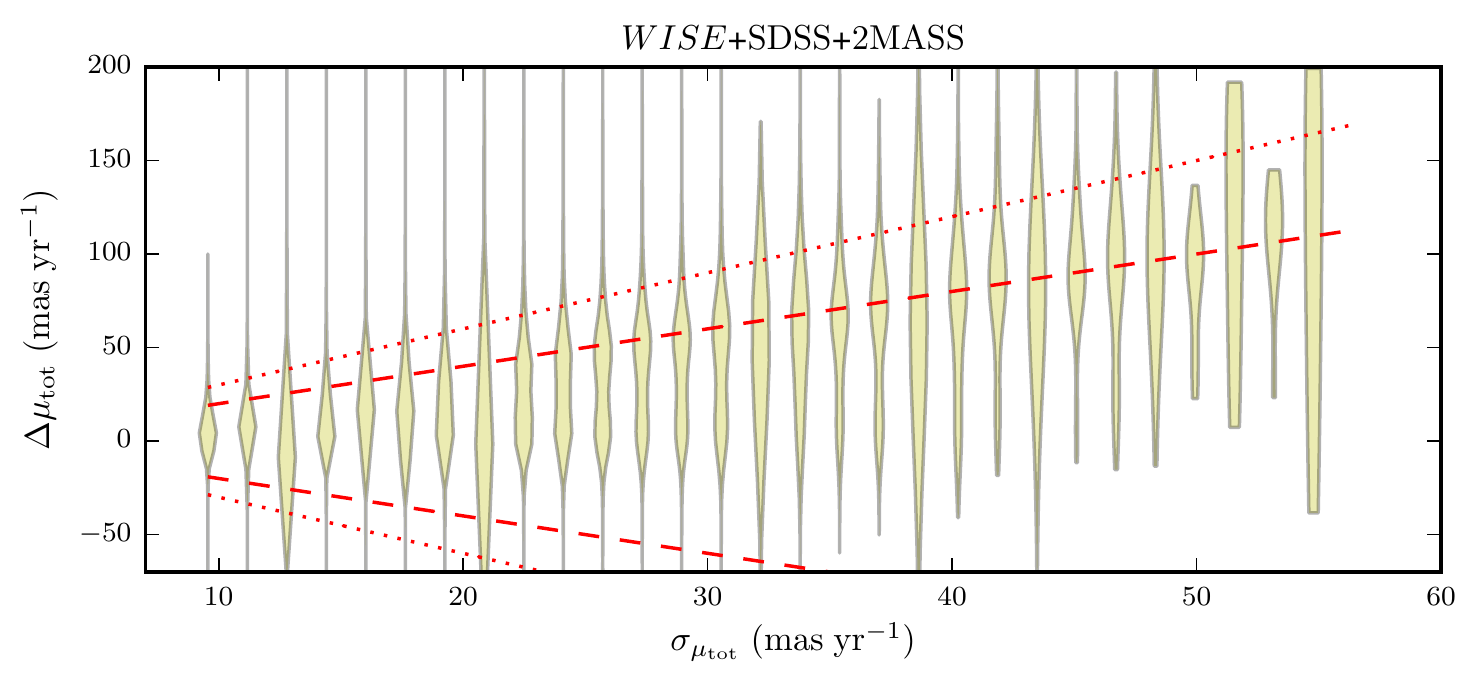}
 \includegraphics[width=.8\textwidth]{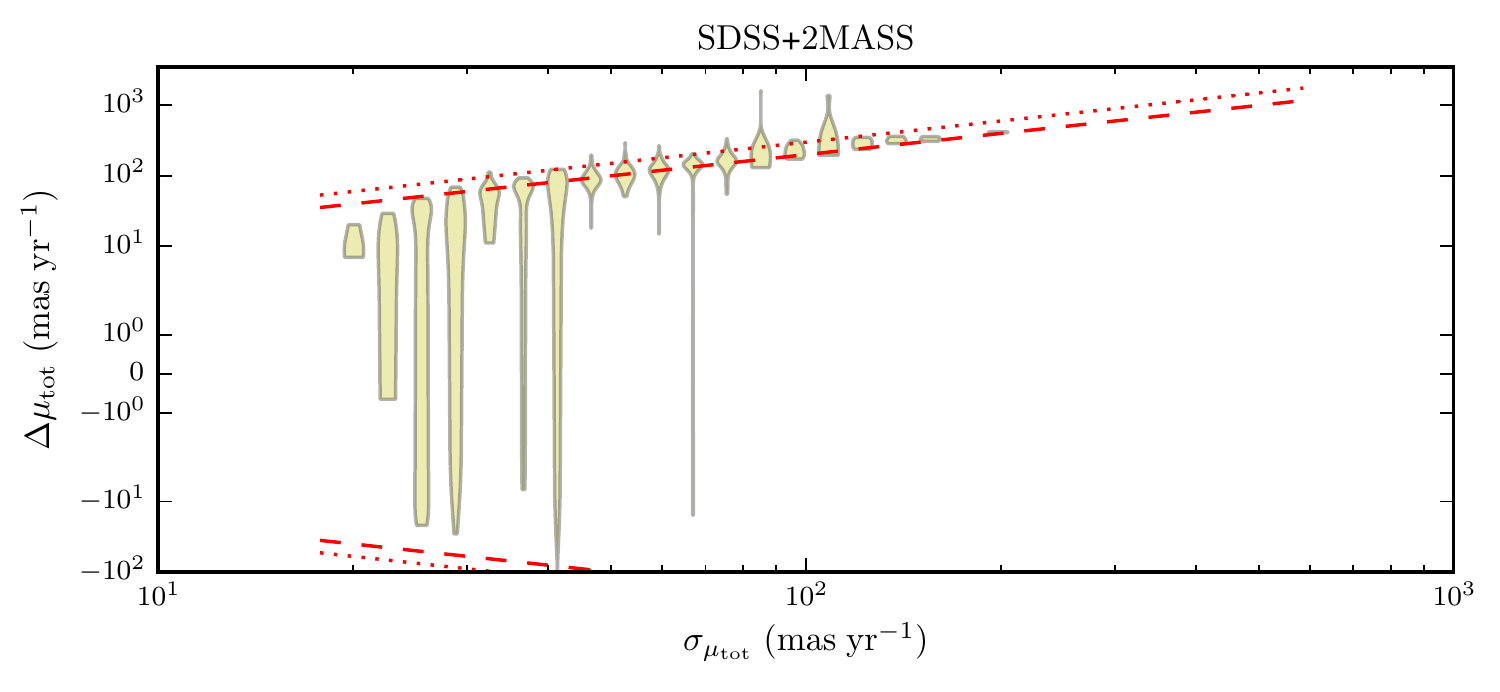}
 \includegraphics[width=.8\textwidth]{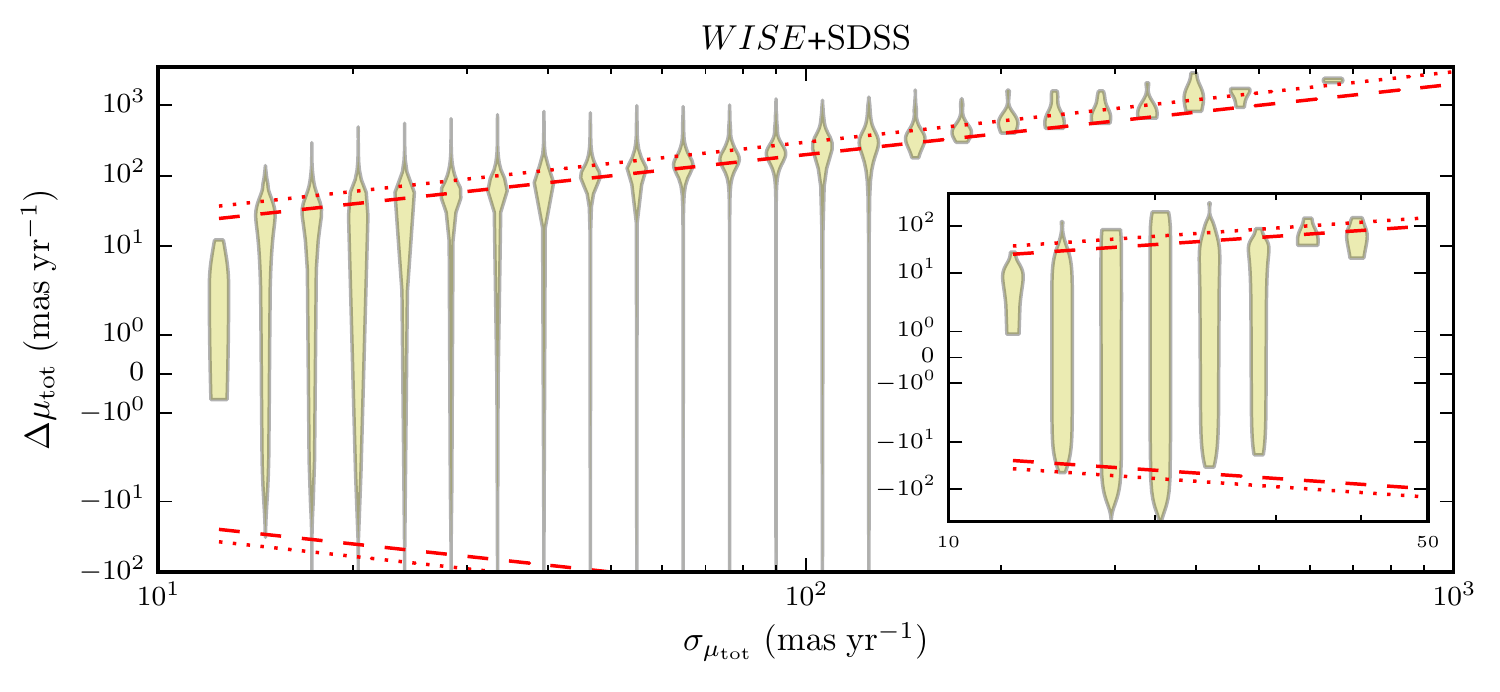}
\caption{Same as Figure~\ref{fig:M04reserr}, but for M14 matches. \textit{Top}: The \textit{WISE}+SDSS+2MASS stars tend to have more scatter in their residuals than the M04 stars, however, the majority are in 3$\sigma$ agreement with M14. \textit{Middle}: The SDSS+2MASS stars tend to be in better agreement for smaller errors, similar to M04 matches. \textit{Bottom}: The \textit{WISE}+SDSS stars also tend to be in better agreement for smaller errors, similar to M04 matches. The inset plot shows only stars with $r-z > 3$, which are typically more reliable.
\label{fig:M14reserr}}
\end{figure*}

	We again explored the 2$\sigma$ and 3$\sigma$ agreement between both proper motion components for M14 and our catalog (see Table~\ref{tbl:PMagreement}). Agreements between our \textit{WISE}+SDSS+2MASS measurements and M14 are similar to those found with M04, except for the \textit{WISE}+SDSS stars which are in better agreement than those found against M04 with a larger number of stars (261,141). Agreement is also better for redder stars (stars with $r-z > 3$), the majority of which were fainter than equation~(\ref{cut1}) from Section~\ref{M04}. This shows that our \textit{WISE}+SDSS sources are typically too faint to be have reliable proper motions in M04. In general, our proper motions across all three subsamples appear to be reliable (to within our uncertainties) up to our magnitude limit of $r = 22$, with redder sources having higher reliability. Figure~\ref{fig:M14reserr} shows how proper motion agreement scales with proper motion error (similar to Figure~\ref{fig:M04reserr}). We find slightly better agreement for stars with larger proper motion errors than we did for M04. As stated above, cutting on proper motion errors can yield higher reliability. 
	
	Our comparison with LSPM shows us that our catalog is reliable for the brightest and faster moving stars. The majority of stars with discrepant proper motions were found to have more reliable measurements in our catalog. Comparisons with M04 and M14 show that our catalog tends to be more reliable for brighter sources, but reliability remains high for fainter, redder stars. Rather than make further cuts on our catalog, we summarize our findings and use the results of our comparisons to suggest selection criteria to obtain a clean sample in Section~\ref{movers}.

\section{Augmenting our Proper Motions with SDSS+USNO-B}\label{usno}

	Our combined \textit{WISE}+SDSS+2MASS proper motions are mostly sensitive to faster moving ($\mu_\mathrm{tot} > 20$ \masyear) and redder stars. To make the most complete set of proper motions and to ensure the highest completeness for low-mass stars within the SDSS photometric sample, we require a baseline more sensitive to slower moving (tangentially) stars, or disk stars. The ideal remedy is M04, which has a higher precision ($\sigma_{\mu_\mathrm{tot}} \lesssim 4$ \masyear), but is more sensitive to brighter sources (see Section~\ref{reliability}). To augment our derived proper motions with those from M04, we must first choose criteria that select the most reliable proper motions from M04. A number of different criteria have been proposed for selecting a ``clean" sample of SDSS+USNO-B stars \citep[e.g.,][]{kilic:2006:582, dhital:2010:2566, dong:2011:116, west:2011:97}. We performed an in-depth exploration into the effect of different criteria on selecting a clean sample.
	
	The most important parameters within the M04 catalog for our investigation were (taken from SDSS SkyServer\footnote{\url{http://skyserver.sdss.org/dr12/en/help/browser/browser.aspx\#&&history=description+ProperMotions+U}} \textsc{ProperMotions} CasJobs table):
\begin{enumerate}
\item \textsc{match}, the number of objects in USNO-B that matched this object within a 1\arcsec\ radius. If negative, then the nearest matching USNO-B object itself matched more than 1 SDSS object.
\item \textsc{sigRA} and \textsc{sigDEC}, the RMS residuals for the proper motion fit (in R.A. and Dec., respectively).
\item \textsc{nFit}, the number of detections used in the fit including the SDSS detection (thus, the number of plates the object was detected on in USNO-B plus one).
\item \textsc{O} and \textsc{J}, the recalibrated USNO-B $O$ and $J$ magnitudes, respectively, recalibrated to SDSS $g$.
\end{enumerate}
We chose to keep the criterion \textsc{dist22} $> 7$ since this criterion is common among all methods for selecting a ``clean" sample. We explored each of the criterion outlined in the aforementioned studies, seeing how proper motion agreement changed with each criteria.

\begin{figure*}
\centering
 \includegraphics{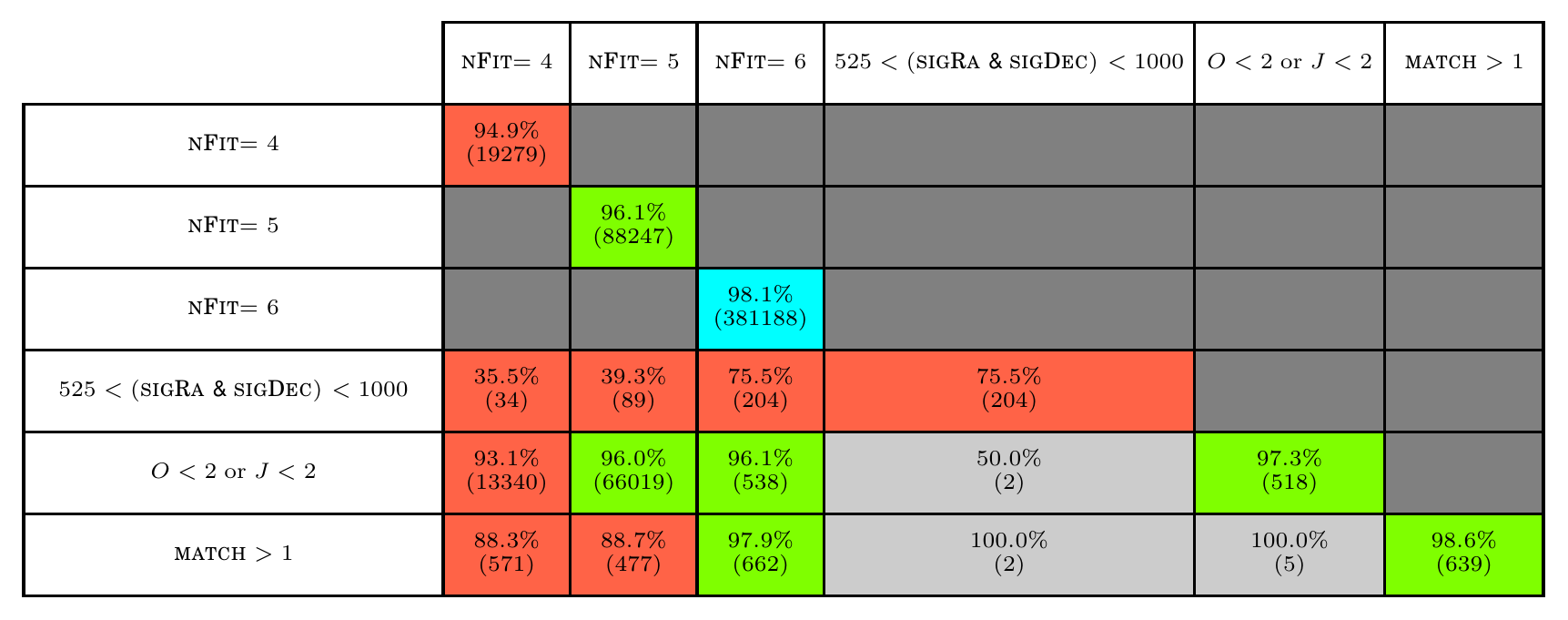}
\caption{Table showing the percentage of stars that had a 2$\sigma$ agreement in both proper motion components ($\alpha$ and $\delta$) between our \textit{WISE}+SDSS+2MASS proper motions and those from M04, as a function of M04 parameters. The number of stars that met each criteria are listed below the percentage in parenthesis. Colors represent our limits for selecting: 1) ``good" criteria (green); 2) ``bad" criteria (red); 3) baseline, or starting, criteria (blue); and 4) criteria that did not include enough stars to be meaningful (light gray). Default criteria, unless changed, were: 1) \textsc{nFit} $ = 6$; 2) \textsc{sigRA} \& \textsc{sigDEC} $< 525$; and 3) \textsc{match} $= 1$.
\label{fig:table}}
\end{figure*}

	To select a clean sample, we chose a subset of our color-magnitude space where the absolute residuals with M04 were typically $< 10$ \masyear. The region we selected from is shown in Figure~\ref{fig:colormag}. In this region, we had 98\% agreement between M04 and our \textit{WISE}+SDSS+2MASS proper motions at 2$\sigma$ (with 381,188 stars). Next, we performed a match between our catalog to M04 with no requirement on any of the above parameters. This match gave us 495,385 stars within the aforementioned color-magnitude space. We performed a number of tests, using criteria recommended in the various papers listed above, and compared the 2$\sigma$ agreement between ours and the M04 proper motion components. The results of our comparisons are summarized in Figure~\ref{fig:table}, and our recommended criteria for selecting high reliability candidates is:

\begin{enumerate}
\item \textsc{dist22} $> 7$ and, 
\item \textsc{sigRA} $< 525$ and \textsc{sigDEC} $< 525$ and, 
\item (\textsc{nFit} $= 6$ and \textsc{match} $\geq 1$) or 
\item (\textsc{nFit} $= 5$ and \textsc{match} $=1$ and [$O < 2$ or $J < 2$]).
\end{enumerate}
Using these criteria should yield proper motions with more than 95\% confidence. Another optional cut is to remove (redder) objects with $r-z > 3$, which typically have less reliable proper motions, however, these objects make up only a small fraction of the objects within M04 ($< 0.2\%$).
	
	We chose to supplement our proper motions with those from M04, applying all of the above criteria outlined, and also requiring a minimum total proper motion greater than two times the combined total proper motion error ($\mu_{\rm tot} > 2\sigma_{\mu_{\rm tot}}$). This gave us 6,620,838 stars, 5,216,855 of which we did not have a prior measurement for. The final number of stars in our catalog for each subsample is listed in Table~\ref{tbl:dbits}.

\section{The M\lowercase{o}V\lowercase{e}RS Catalog}\label{movers}

\begin{figure}
\centering
 \includegraphics{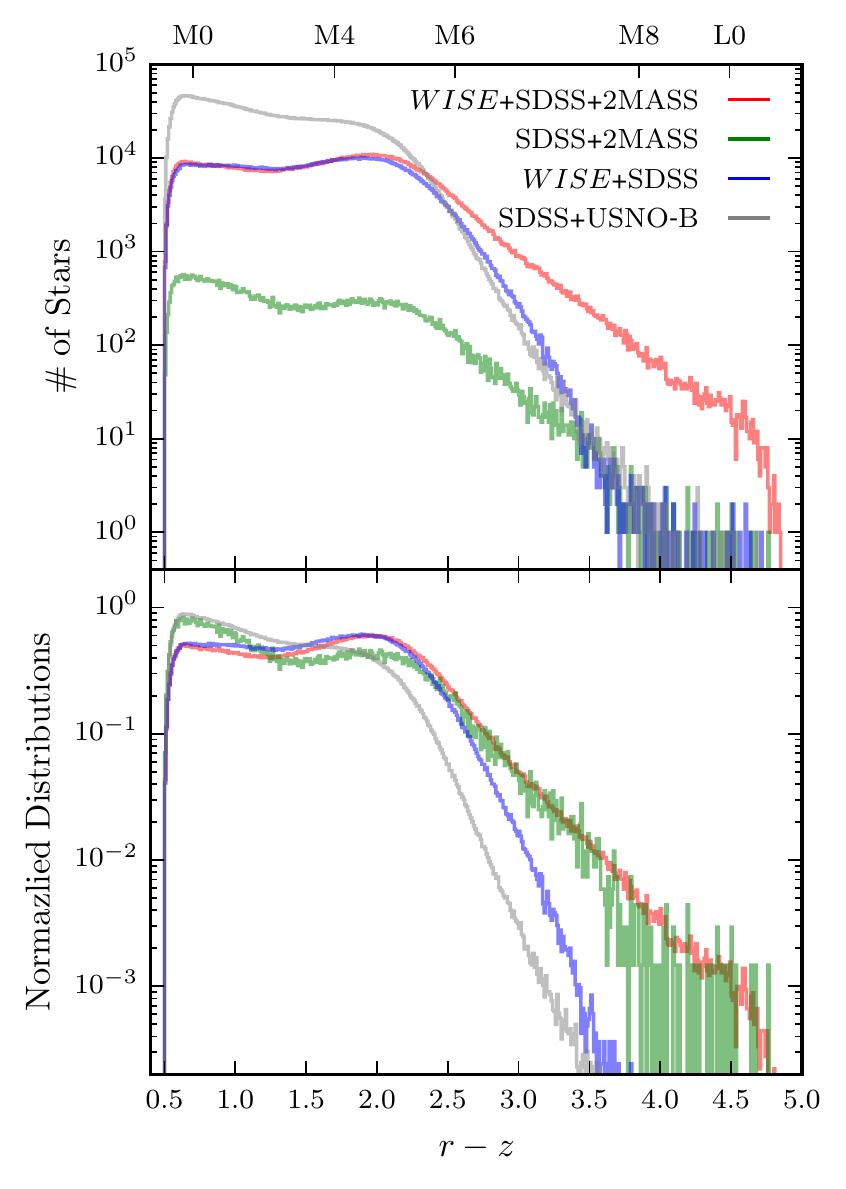}
\caption{$r-z$ distributions for each subsample in our catalog. Approximate spectral types taken from \citet{hawley:2002:3409} and \citet{bochanski:2007:531}. M04 and our SDSS+2MASS distributions typically include more bluer stars, while proper motions measured with \textit{WISE} typically include more redder stars.
\label{fig:rz}}
\end{figure}

\begin{figure*}
\centering
 \includegraphics{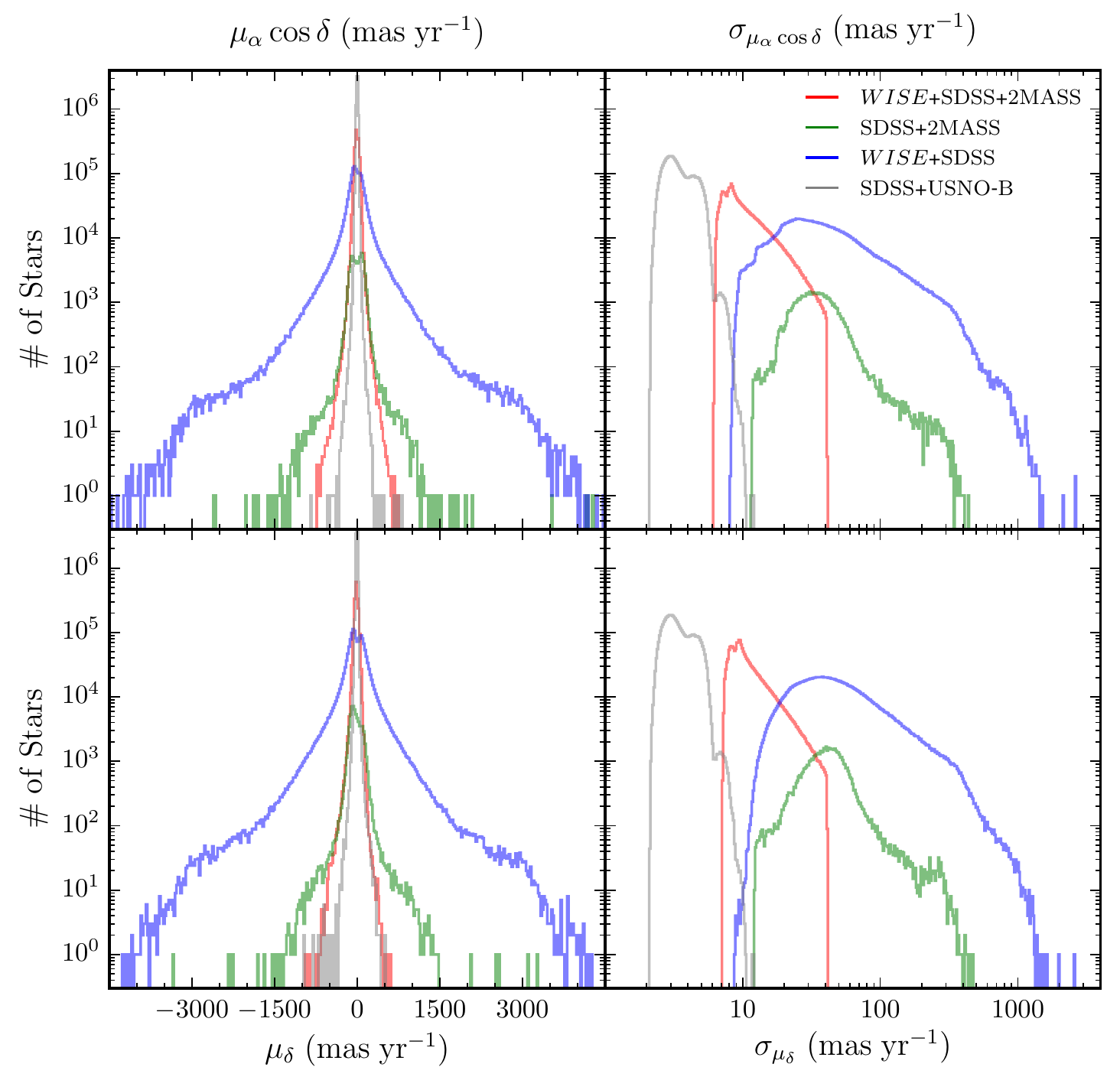}
\caption{Proper motion and proper motion error distributions for our catalog. The vast majority of our catalog is comprised of stars with $\mu_{\rm tot} < 1000$ \masyear\ (99.5\% of the entire catalog).
\label{fig:distpms}}
\end{figure*}

	Our goal was not to construct a ``complete" proper motion catalog, rather, we have attempted to build a catalog of low-mass stars with high-fidelity proper motion measurements. We expect our catalog to be more ``complete" at the red end of the main-sequence than most previous proper motion catalogs due to the use of deeper surveys, specifically in the IR. To illustrate this point, we show the $r-z$ color distribution for each of our subsamples (e.g., \textit{WISE}+SDSS+2MASS or SDSS+USNO-B) in Figure~\ref{fig:rz}. Our \textit{WISE}+SDSS+2MASS baseline contains a large number of star with $r-z > 2.5$, and our \textit{WISE}+SDSS subsample also contains more stars with $r-z > 2.5$ than our M04 stars, which peak at a spectral type of $\sim$M4. Our newly computed proper motions will identify many new very-low-mass stars with $r-z > 2.5$ (spectral types later than $\sim$M6). The schema for our catalog can be found in Appendix~\ref{schema}. Proper motions and errors for each of the subsamples in our catalog are shown in Figure~\ref{fig:distpms}. The extremely wide wings for our \textit{WISE}+SDSS and SDSS+2MASS matches are due primarily to stars with short time baselines ($<$ 3 years), which typically overestimate proper motions.

	A small number of our stars have high proper motions ($\mu_{\rm tot} > 1000$ \masyear; $\sim$0.5\%). Some of these stars are due to our large search radius (6\arcsec), which may pull in neighboring objects. Our color selection criteria should remove a number of these, however, in crowded fields (e.g., within the Galactic plane), these objects are more prominent. Rather than remove fast moving objects, since many of them will be true detections, we recommend a conservative cut to remove potential outliers is to eliminate stars near the Galactic plane (e.g., $|b| < 20^\circ$).
	
	Another potential cut is to remove objects that have appeared to move a distance close to our search radius over the time baseline (e.g., objects that have moved $\sim$6\arcsec\ within a 1 year period). We plan to determine the validity of these high proper motions sources in a future study. Also, selecting stars that either satisfy equation~(\ref{cut1}) or (\ref{cut2}) can yield higher reliability proper motions, but is biased towards selecting stars with all three epochs. Lastly, selecting stars with smaller proper motion errors ($\sigma_{\mu_\mathrm{tot}} \lesssim 60$ \masyear), specifically for the \textit{WISE}+SDSS and SDSS+2MASS subsamples can increase reliability. The M04

\section{Discussion}\label{discussion}

\begin{figure*}
\centering
 \includegraphics{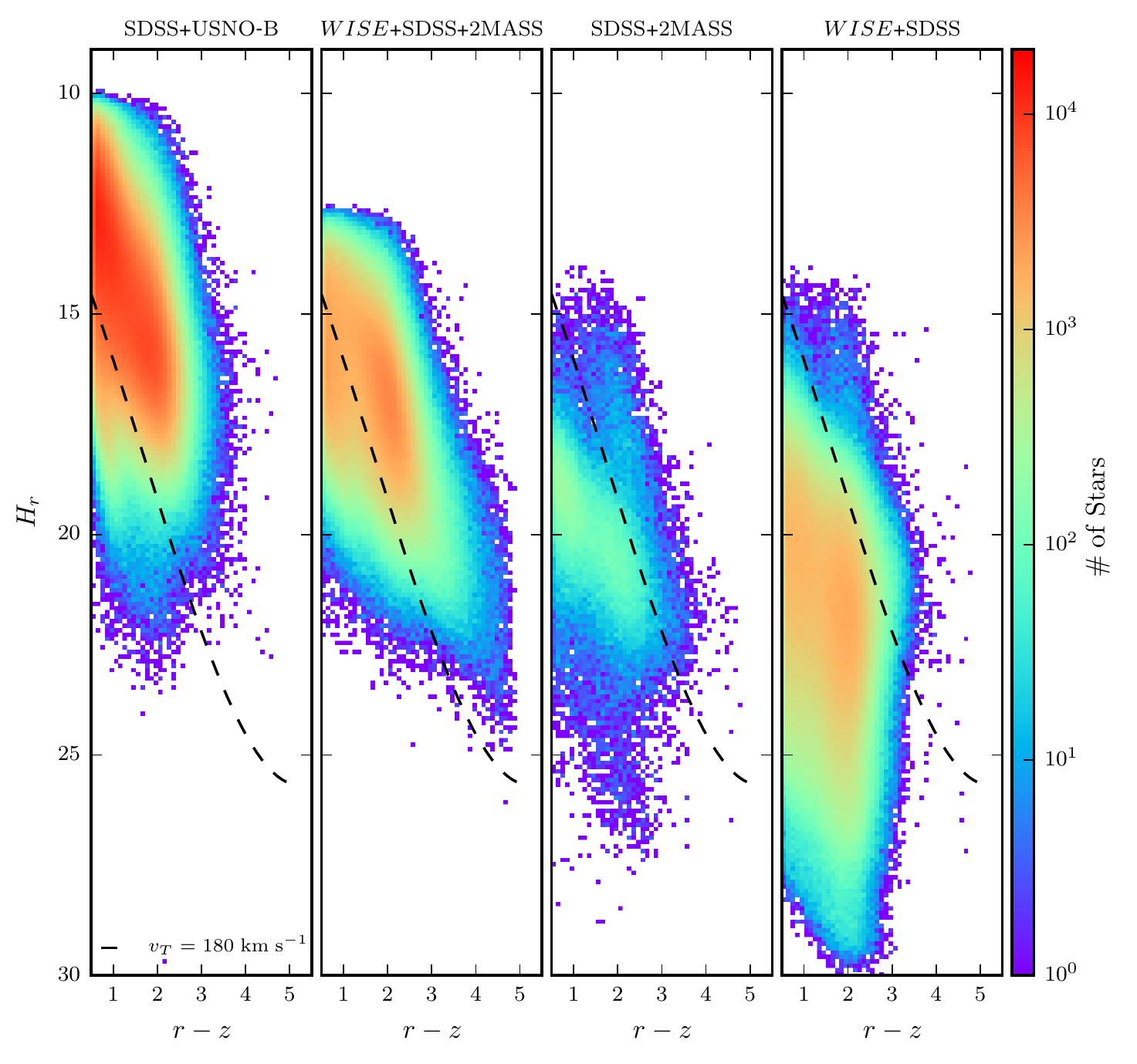}
\caption{Reduced proper motion diagrams for the four sub-samples included in our catalog, each bin is (0.1 mags)$^{2}$. The dashed line represent a tangential velocity of 180 km s$^{-1}$, which separates disk stars from halo stars \citep{sesar:2008:1244}. Our catalog provides one of the largest samples of high-confidence subdwarfs (stars below the dashed line). The M04 source at $r-z \approx 2$, $H_r \approx 30$ appears to have a spurious proper motion (as observed in archived images).
\label{fig:rpm}}
\end{figure*}

	This paper presents an improved proper motion catalog for low-mass stars, specifically, completing the red end of the low-mass sequence. Based off our color selection, this catalog spans from late K dwarfs to early L dwarfs. In particular, this catalog will allow unprecedented studies of subdwarfs and potential halo stars. To show the relative populations of stars within our catalog, we include reduced proper motions \citep[RPMs;][]{luyten:1922:135}, given as
\begin{equation}
H_r = r + 5 + 5 \log \mu = M_r - 3.25 + 5 \log v_T,
\end{equation}
where $\mu$ is the total proper motion in arcsec yr$^{-1}$, and $v_T$ is the heliocentric tangential velocity in km s$^{-1}$ given by $v_T = 4.74 \times (\mu\cdot d)$, where $d$ is the distance in parsecs. RPM diagrams for each of our four subsamples are shown in Figure~\ref{fig:rpm}. To segregate different kinematic groups, we have drawn a line at $v_T = 180$ km s$^{-1}$, the approximate limit which differentiates disk stars from subdwarfs \citep{sesar:2008:1244}. Although subdwarfs may scatter above the 180 km s$^{-1}$ line, disk dwarfs are not typically found below the 180 km s$^{-1}$ line \citep{dhital:2010:2566}. 

	The SDSS+USNO-B stars are primarily earlier type disk stars. Our \textit{WISE}+SDSS+2MASS stars are also primarily disk dwarfs, but peak at redder colors, and are more complete (and reliable) at the reddest end of the main-sequence. Our SDSS+2MASS and \textit{WISE}+SDSS baselines appear to probe bluer and redder subdwarf populations, respectively.
	
	We have identified a number of science questions which can be investigated with this catalog:
	\begin{enumerate}
	\item Identify SDSS spectroscopic high-proper motion low-mass stars to find hypervelocity candidates (Favia et al., submitted.).
	\item Identify extremely low-mass, common proper motion binaries.
	\item Investigate Galactic kinematics for the lowest-mass main-sequence members.
	\item Identify low-mass field stars with infrared excesses.
	\item Confirm pervious catalogs of wide-binaries \citep[e.g.,][]{dhital:2015:57}.
	\end{enumerate}
	With current and future efforts multi-epoch surveys, such as \textit{Gaia} and LSST, this catalog will also prove invaluable as a calibrator for the reddest stellar populations within these surveys.
	
\subsection{Common Proper Motion Binaries: Investigating the SloWPoKES-II Catalog}

\begin{figure*}
\centering
 \includegraphics{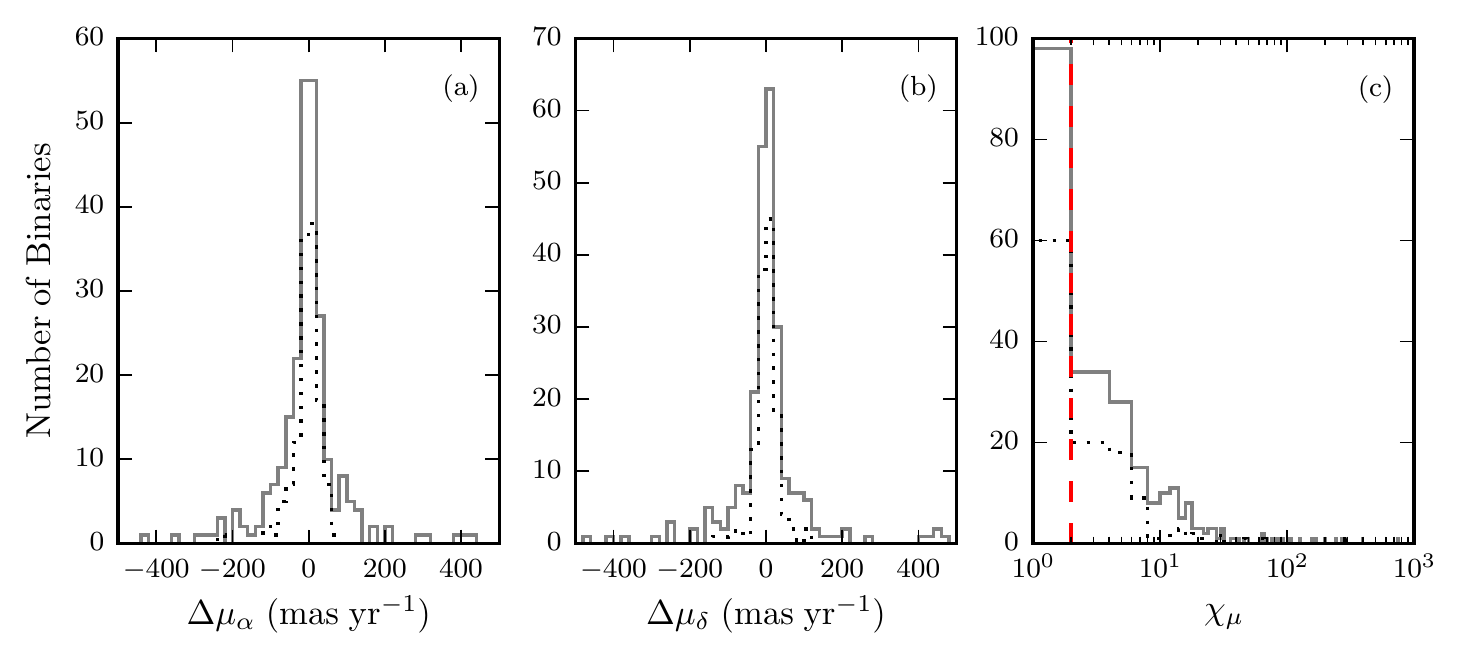}
\caption{Distributions for proper motion residuals between the primary and secondary components for the stars matched between our MoVeRS catalog and SLoWPoKES-II. The solid line denotes all pairs matched, and the dotted line denotes only stars matched with all three epochs (i.e. \textit{WISE}+SDSS+2MASS). Subplots (a) and (b) show the proper motion differences between the primary and secondary in $\alpha$ and $\delta$, respectively. Both distributions are peaked at zero (dashed lines), showing good agreement between the proper motions. Subplot (c) shows equation~(\ref{eqn:slw}), the quadrature sum of the weighted difference of both proper motion components. The red dotted line denotes a value of 2, the cutoff value for binaries to be include in the original SLoWPoKES catalog.
\label{fig:slw2}}
\end{figure*}

	To demonstrate the utility of MoVeRS, we matched our catalog to the second release of the Sloan Low-mass Wide Pairs of Kinematically Equivalent Stars \citep[SLoWPoKES-II;][]{dhital:2015:57}, a sample of wide binaries identified without proper motions. This catalog is an extension of the study from \citet[SLoWPoKES-I;][]{dhital:2010:2566}, where it was shown that wide-binaries with separations less than $\sim$20\arcsec\ could be identified based off similar distances, but without the need for proper motions. There were 260 matches between the two catalogs. Figure~\ref{fig:slw2} shows the distributions for the proper motion components between the primary and secondary, the distribution of equation (6) from \citet{dhital:2010:2566}, a measure of the weighted difference in both proper motion components given as
\begin{equation}\label{eqn:slw}
\chi_\mu = \left(\frac{\Delta\mu_\alpha}{\sigma_{\Delta\mu_\alpha}}\right)^2 + \left(\frac{\Delta\mu_\delta}{\sigma_{\Delta\mu_\delta}}\right)^2.
\end{equation}
The fraction of matches that met the criteria of reliable proper motions binaries \citep[$\chi_\mu \leq 2$;][]{dhital:2010:2566} was 38\%. If we restrict this analysis to pairs that have proper motions with all three epochs (i.e. \textit{WISE}+SDSS+2MASS; 127 pairs), this fraction increase to 47\%.

\begin{figure}
\centering
 \includegraphics{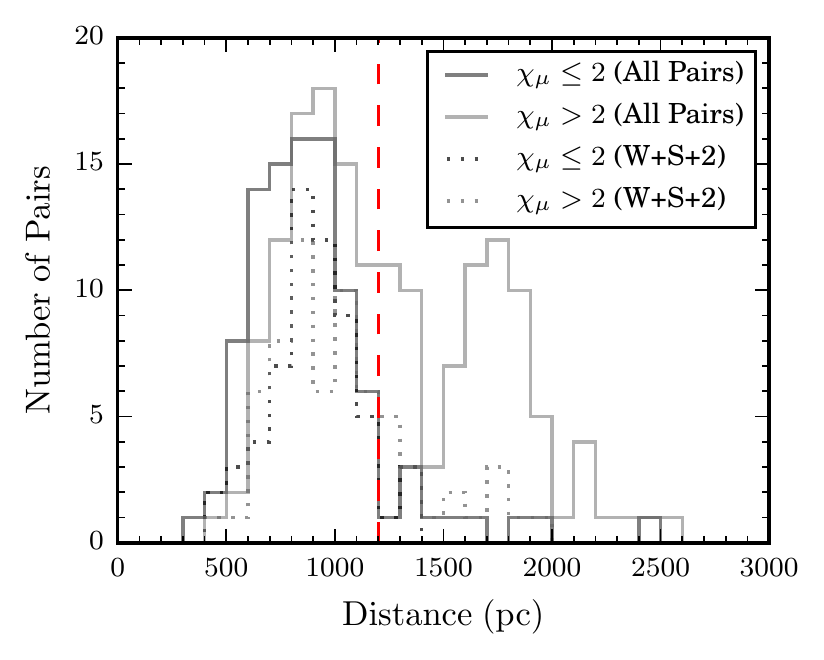}
\caption{Average distance distributions for pairs from the SLoWPoKES-II catalog. Dark gray lines correspond to high-fidelity pairs ($\chi_\mu \leq 2$; equation~\ref{eqn:slw}), and light grey to low-fidelity ($\chi_\mu > 2$) pairs. Dotted lines correspond to all pairs found within MoVeRS, while solid lines correspond to only pairs where both components had three epochs (i.e. \textit{WISE}+SDSS+2MASS). Pairs with an average distance $> 1200$ pc (dashed line) tend to have less similar proper motions (and hence lower fidelity) than closer pairs. The majority of pairs within SLoWPoKES-II have angular separations $< 8\arcsec$, so we expect many of the pairs to have high-reliability.
\label{fig:slw2dist}}
\end{figure}

	The original SLoWPoKES catalog did not have a formal constraint on average distance due to the fact that the range of spectral types (mid-K to mid-M) made the catalog sensitivity a function of color and distance. It was found that reliability was higher for stars with $d \lesssim 1200$ pc, due to smaller photometric distance uncertainties. Since SLoWPoKES-II has no constraint on average distance, we chose to explore how pair fidelity corresponded with distance. Figure~\ref{fig:slw2dist} shows the distribution of both high-fidelity pairs ($\chi_\mu \leq 2$) and low-fidelity pairs ($\chi_\mu > 2$). High-fidelity pairs are primarily found within 1200 pc, with fidelity becoming noticeably worse at larger distances. 
	
\begin{figure*}
\centering
 \includegraphics{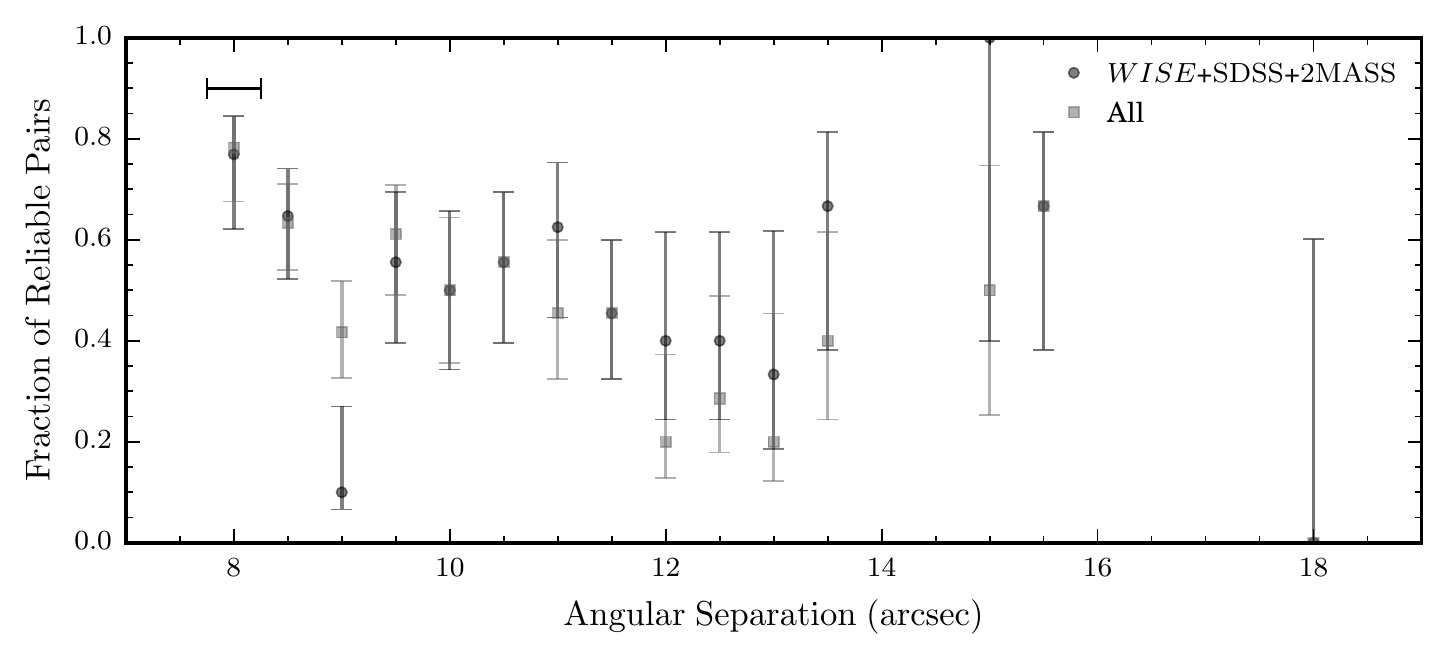}
\caption{Reliability fraction of pairs (with reliable pairs defined as pairs with $\chi_\mu \leq 2$) as a function of angular separation between components. The bin size is shown in the top left corner. Only pairs with average distances $\leq 1200$ pc were used. Reliability fractions were calculated for all matched pairs (light gray square), and pairs where both components had \textit{WISE}+SDSS+2MASS measurements (dark gray circles). We did not find any reliable pairs with angular separations $> 15$\arcsec. Reliability increases for smaller angular separations, unfortunately we cannot probe the majority of the SLoWPoKES-II pairs, which have angular separations $< 8$\arcsec.
\label{fig:slw2sep}}
\end{figure*}

	We chose to investigate how angular separation between the primary and secondary component affected the likelihood of pairs having common proper motions. From the results of \citet{dhital:2010:2566} and \citet{dhital:2015:57}, we expect closer pairs to have a higher probability of being true binaries. However, we only have the ability to probe pairs with separations $\gtrsim 8\arcsec$, except for a small subset of objects matched in SDSS+2MASS. To investigate reliability as a function of pair separation, we calculated the fraction of reliable pairs in a moving 1\arcsec bin and computed binomial uncertainties, using only pairs with an average distance $\leq 1200$ pc since reliability was shown to decrease significantly past this point. Our results are shown in Figure~\ref{fig:slw2sep}; as expected, proper motion reliability decreases with angular separation; for sources with angular separations $> 15\arcsec$, we find reliability drops to zero. Due to the small sample size of our matched pairs, we do not attempt to make any claims on the overall reliability of SLoWPoKES-II; this exercise was simply used to show the applicability of our catalog. From our initial results, coupled with the fact that the distribution of angular separations in the SLoWPoKES-II catalog peaks at separations $< 8\arcsec$, we expect many of the pairs to be \textit{bona-fide} binaries. Choosing pairs with smaller average distances and smaller angular separations can help improve the fidelity of the sample. 

\section{Summary}\label{summary}

	We have created a catalog containing \stars\ proper motion verified photometric low-mass stars. Proper motions were computed using the \textit{WISE}, SDSS, and 2MASS surveys (\starsUs\ stars), and augmented with proper motions from SDSS+USNO-B \citep[\starsMunn\ stars;][]{munn:2004:3034,munn:2008:895}. All stars were required to have a total proper motion greater than twice the uncertainty in their measurement, thus ensuring high-fidelity main-sequence stars. The estimated precision of our catalog is $\sim$10 \masyear\ for our \textit{WISE}+SDSS+2MASS sources, and $\sim$40 \masyear\ for our SDSS+2MASS and \textit{WISE}+SDSS sources, primarily due to shorter time baselines.

	Comparison against the high proper motions stars from LSPM suggests good agreement for high proper motion stars in our catalog in all three subsamples (e.g., \textit{WISE}+SDSS or SDSS+2MASS). For our subsamples, agreement with LSPM at the 2$\sigma$ level was 98\%, 97\% and 96\% for \textit{WISE}+SDSS+2MASS, SDSS+2MASS, and \textit{WISE}+SDSS, respectively (these all increase to 99\% at the 3$\sigma$ level). We further compared our proper motions to SDSS+USNO-B measurements and the deeper proper motion catalog released by \citet{munn:2014:132}. In both cases, our proper motion precision is strongly correlated with apparent magnitude and color, diminishing for bluer and fainter sources.
	
	The utility of this catalog will be in the vast number of motion verified low-mass stars it contains, and its high reliability, specifically for the reddest and lowest-mass members of the catalog. We expect the red end of this catalog to surpass the limits of \textit{Gaia}. Our catalog is available through SDSS CasJobs and VizieR.

\acknowledgments
	The authors would first like to thank the anonymous referee for their extremely helpful comments which greatly improved the quality of this study. The authors would like to thank Ani Thankar for support running long queries on SDSS CasJobs on the behalf of this study, and for help uploading the catalog to CasJobs. The authors would also like to thank Benjamin Alan Weaver and Vandana Desai for their help with understanding the SDSS and 2MASS data, respectively. C.A.T. would like to thank Dylan Morgan, Julie Skinner, and Brandon Harrison for many helpful discussions. C.A.T. would like to acknowledge the Ford Foundation for financial support. A.A.W acknowledges funding from NSF grants AST-1109273 and AST-1255568. A.A.W. and C.A.T. also acknowledge the support of the Research Corporation for Science Advancement's Cottrell Scholarship.

	Funding for SDSS-III has been provided by the Alfred P. Sloan Foundation, the Participating Institutions, the National Science Foundation, and the U.S. Department of Energy Office of Science. The SDSS-III web site is \url{http://www.sdss3.org/.}

SDSS-III is managed by the Astrophysical Research Consortium for the Participating Institutions of the SDSS-III Collaboration including the University of Arizona, the Brazilian Participation Group, Brookhaven National Laboratory, Carnegie Mellon University, University of Florida, the French Participation Group, the German Participation Group, Harvard University, the Instituto de Astrofisica de Canarias, the Michigan State/Notre Dame/JINA Participation Group, Johns Hopkins University, Lawrence Berkeley National Laboratory, Max Planck Institute for Astrophysics, Max Planck Institute for Extraterrestrial Physics, New Mexico State University, New York University, Ohio State University, Pennsylvania State University, University of Portsmouth, Princeton University, the Spanish Participation Group, University of Tokyo, University of Utah, Vanderbilt University, University of Virginia, University of Washington, and Yale University.
 
	This publication makes use of data products from the Two Micron All Sky Survey, which is a joint project of the University of Massachusetts and the Infrared Processing and Analysis Center/California Institute of Technology, funded by the National Aeronautics and Space Administration and the National Science Foundation. 
	
	This publication also makes use of data products from the \textit{Wide-field Infrared Survey Explorer}, which is a joint project of the University of California, Los Angeles, and the Jet Propulsion Laboratory/California Institute of Technology, funded by the National Aeronautics and Space Administration. 
		
	This research made use of Astropy, a community-developed core Python package for Astronomy \citep{astropy-collaboration:2013:a33}. Figures in this work were created using the Python based graphics environment Matplotlib \citep{hunter:2007:90}. 
	
	The authors are also pleased to acknowledge that much of the computational work reported on in this paper was performed on the Shared Computing Cluster which is administered by Boston University's Research Computing Services (\url{www.bu.edu/tech/support/research/}). In particular, C.A.T. would like to thank Paul Dalba and Yann Tambouret for helpful discussions regarding optimization of code and parallelization.

\appendix

\section{A.1 Color Selection Polygons}\label{polys}
	Tables~\ref{tbl:poly1}, \ref{tbl:poly2}, and \ref{tbl:poly3} contain the color selection criteria we used to trace the stellar locus and select our initial sample.
	
\begin{deluxetable}{c c c c}
\tabletypesize{\footnotesize}
\tablecolumns{4}
\tablecaption{$i-z$ Color Selection Criteria\label{tbl:poly1}}
\tablehead{
\colhead{$r-i$} & \colhead{$i-z$} & \colhead{\# of Stars} & \colhead{Bin Size}
}
\startdata
0.200 & $0.367 \pm 0.044$ & 309784 & 0.01 \\
0.205 & $0.367 \pm 0.044$ & 309784 & 0.01 \\
0.215 & $0.379 \pm 0.048$ & 318938 & 0.01 \\
... & ... & ... & ... \\
2.010 & $2.685 \pm 0.210$ & 8 & 0.2 \\
2.020 & $2.685 \pm 0.210$ & 8 & 0.2 \\
2.030 & $2.685 \pm 0.210$ & 8 & 0.2 
\enddata
\tablecomments{Table~\ref{tbl:poly1} is published in its entirety in the electronic edition of AJ, a portion is shown here for guidance regarding its form and content.}
\end{deluxetable}

\begin{deluxetable}{c c c c}
\tabletypesize{\footnotesize}
\tablecolumns{4}
\tablecaption{$z-J$ Color Selection Criteria\label{tbl:poly2}}
\tablehead{
\colhead{$r-z$} & \colhead{$z-J$} & \colhead{\# of Stars} & \colhead{Bin Size}
}
\startdata
0.500 & $1.030 \pm 0.056$ & 1384 & 0.01 \\
0.505 & $1.030 \pm 0.056$ & 1384 & 0.01 \\
0.515 & $1.033 \pm 0.054$ & 3861 & 0.01 \\
... & ... & ... & ... \\
4.900 & $2.076 \pm 0.049$ & 3 & 0.2 \\
4.910 & $2.044 \pm 0.024$ & 2 & 0.2 \\
4.920 & $2.044 \pm 0.024$ & 2 & 0.2 
\enddata
\tablecomments{Table~\ref{tbl:poly2} is published in its entirety in the electronic edition of AJ, a portion is shown here for guidance regarding its form and content.}
\end{deluxetable}

\begin{deluxetable}{c c c c}
\tabletypesize{\footnotesize}
\tablecolumns{4}
\tablecaption{$z-W1$ Color Selection Criteria\label{tbl:poly3}}
\tablehead{
\colhead{$r-z$} & \colhead{$z-W1$} & \colhead{\# of Stars} & \colhead{Bin Size}
}
\startdata
0.500 & $1.712 \pm 0.096$ & 4889 & 0.01 \\
0.505 & $1.712 \pm 0.096$ & 4889 & 0.01 \\
0.515 & $1.724 \pm 0.096$ & 13629 & 0.01 \\
... & ... & ... & ... \\
4.900 & $3.429 \pm 0.082$ & 3 & 0.2 \\
4.910 & $3.373 \pm 0.027$ & 2 & 0.2 \\
4.920 & $3.373 \pm 0.027$ & 2 & 0.2 
\enddata
\tablecomments{Table~\ref{tbl:poly3} is published in its entirety in the electronic edition of AJ, a portion is shown here for guidance regarding its form and content.}
\end{deluxetable}

\section{A.2 Querying the Catalog}\label{query}

	Our Motion Verified Red Stars (MoVeRS) catalog is available through SDSS CasJobs\footnote{\url{http://skyserver.sdss.org/casjobs/}} and VizieR\footnote{\url{http://vizier.u-strasbg.fr/viz-bin/VizieR}}. To access our catalog through CasJobs, please refer to the documentation for accessing public tables. The following is an example SQL query for accessing our table within the DR10 context to return SDSS positions, $rizJHKW1$ photometry, proper motions, and proper motion errors for stars with: 1) total proper motions less then 500 \masyear; 2) total proper motion errors less than 20 \masyear; 3) $r-z > 2.5$; and 4) $18 < r < 21$:
	
\begin{verbatim}
   SELECT 
       p.sdss_ra, p.sdss_dec, p.rmag, p.imag, p.zmag, p.jmag, p.hmag, p.kmag, 
       p.w1mpro, p.pmra, p.pmdec, p.pmra_toterr, p.pmdec_toterr
   FROM public.LowMassPM.MoVeRS p
   WHERE 
       p.pmra * p.pmra + p.pmdec * p.pmdec < 500 * 500 AND
       p.pmra_toterr * p.pmra_toterr + p.pmdec_toterr * p.pmdec_toterr < 20 * 20 AND
       p.rmag - p.zmag > 2.5 AND
       p.rmag BETWEEN 18 and 21
\end{verbatim}

\section{A.3 MoVeRS Catalog Schema}\label{schema}

\LongTables
\begin{deluxetable*}{llll}
\tabletypesize{\footnotesize}
\tablecolumns{4}
\tablecaption{Catalog Schema\label{tbl:schema}}
\tablehead{
Field Name & Format & Units & Description
}
\startdata
SDSS\_OBJID				& int64				&				& SDSS DR8+ Object ID \\
SDSS\_RA 				& float64				& degrees			& SDSS R.A.\\
SDSS\_DEC 				& float64				& degrees			& SDSS Decl.\\
SDSS\_RAERR 			& float32				& degrees			& SDSS R.A. error (in proper units, i.e. $\Delta\alpha \cos \delta$)\\
SDSS\_DECERR 			& float32				& degrees			& SDSS Decl. error\\
SDSS\_MJD		 		& float32				& days			& SDSS $r$-band modified Julian date\\
UMAG			 		& float32				& mags			& SDSS $u$-band PSF magnitude\\
GMAG			 		& float32				& mags			& SDSS $g$-band PSF magnitude\\
RMAG		 			& float32				& mags			& SDSS $r$-band PSF magnitude\\
IMAG		 			& float32				& mags			& SDSS $i$-band PSF magnitude\\
ZMAG			 		& float32				& mags			& SDSS $z$-band PSF magnitude\\
UMAG\_ERR		 		& float32				& mags			& SDSS $u$-band PSF magnitude error\\
GMAG\_ERR	 			& float32				& mags			& SDSS $g$-band PSF magnitude error\\
RMAG\_ERR	 			& float32				& mags			& SDSS $r$-band PSF magnitude error\\
IMAG\_ERR		 		& float32				& mags			& SDSS $i$-band PSF magnitude error\\
ZMAG\_ERR		 		& float32				& mags			& SDSS $z$-band PSF magnitude error\\
2MASS\_RA 				& float32				& degrees			& 2MASS R.A.\\
2MASS\_DEC 				& float32				& degrees			& 2MASS Decl.\\
2MASS\_RAERR 			& float32				& degrees			& 2MASS R.A. error (in proper units, i.e. $\Delta\alpha \cos \delta$)\\
2MASS\_DECERR 			& float32				& degrees			& 2MASS Decl. error\\
2MASS\_MJD	 			& float32				& days			& 2MASS modified Julian date\\
2MASS\_PH\_QUAL	 		& 3 character string		& 				& 2MASS photometric quality flag\\
2MASS\_RD\_FLG	 		& 3 character string		& 				& 2MASS read flag\\
2MASS\_BL\_FLG	 		& 3 character string		& 				& 2MASS blend flag\\
2MASS\_CC\_FLG	 		& 3 character string		& 				& 2MASS contamination and confusion flag\\
2MASS\_GAL\_CONTAM		& int32				& 				& 2MASS extended source ``contamination" flag\\
JMAG					& float32				& mags			& 2MASS $J$-band PSF magnitude\\
JMAG\_ERR				& float32				& mags			& 2MASS $J$-band PSF corrected magnitude uncertainty\\
JMAG\_ERRTOT			& float32				& mags			& 2MASS $J$-band PSF total magnitude uncertainty\\
JSNR					& float32				& mags			& 2MASS $J$-band SNR\\
HMAG					& float32				& mags			& 2MASS $H$-band PSF magnitude\\
HMAG\_ERR				& float32				& mags			& 2MASS $H$-band PSF corrected magnitude uncertainty\\
HMAG\_ERRTOT			& float32				& mags			& 2MASS $H$-band PSF total magnitude uncertainty\\
HSNR					& float32				& mags			& 2MASS $H$-band SNR\\
KMAG					& float32				& mags			& 2MASS $K_s$-band PSF magnitude\\
KMAG\_ERR				& float32				& mags			& 2MASS $K_s$-band PSF corrected magnitude uncertainty\\
KMAG\_ERRTOT			& float32				& mags			& 2MASS $K_s$-band PSF total magnitude uncertainty\\
KSNR					& float32				& mags			& 2MASS $K_s$-band SNR\\
J\_PSFCHI				& float32				&				& 2MASS $J$-band reduced $\chi^2$ goodness-of-fit for the PSF\\
H\_PSFCHI				& float32				&				& 2MASS $H$-band reduced $\chi^2$ goodness-of-fit for the PSF\\
K\_PSFCHI				& float32				&				& 2MASS $K_s$-band reduced $\chi^2$ goodness-of-fit for the PSF\\
WISE\_RA 				& float32				& degrees			& \textit{WISE} R.A.\\
WISE\_DEC 				& float32				& degrees			& \textit{WISE} Decl.\\
WISE\_RAERR 			& float32				& degrees			& \textit{WISE} R.A. error (in proper units, i.e. $\Delta\alpha \cos \delta$)\\
WISE\_DECERR 			& float32				& degrees			& \textit{WISE} Decl. error\\
WISE\_CC\_FLG	 		& 4 character string		& 				& \textit{WISE} contamination and confusion flag\\
WISE\_EXT\_FLG	 		& int32				& 				& \textit{WISE} extended source flag\\
WISE\_VAR\_FLG	 		& 4 character string		& 				& \textit{WISE} variability flag\\
WISE\_PH\_QUAL	 		& 4 character string		& 				& \textit{WISE} photometric quality flag\\
WISE\_W1MJDMEAN	 	& float32				& days			& \textit{WISE} $W1$-band average modified Julian date\\
W1MJDSIG	 			& float32				& days			& \textit{WISE} MJD uncertainty\tablenotemark{a}\\
W1MPRO	 				& float64				& mags			& \textit{WISE} $W1$-band PSF magnitude\\
W1SIGMPRO		 		& float64				& mags			& \textit{WISE} $W1$-band PSF magnitude uncertainty\\
W1SNR	 				& float64				& 				& \textit{WISE} $W1$-band SNR\\
W1RCHI2			 		& float32				& 				& \textit{WISE} reduced $\chi^2$ goodness-of-fit for the PSF\\
NEAREST\_NEIGHBOR 		& float32				& arcsec			& Distance to nearest SDSS primary object\\
NEAREST\_RMAG	 		& float32				& mags			& SDSS $r$-band PSF magnitude of nearest neighbor\\
NEAREST\_IMAG	 		& float32				& mags			& SDSS $i$-band PSF magnitude of nearest neighbor\\
NEAREST\_ZMAG	 		& float32				& mags			& SDSS $z$-band PSF magnitude of nearest neighbor\\
NEIGHBORS		 		& int32				& 				& Number of SDSS primary objects within 15\arcsec \\
RR1				 		& int32				& 				& Flag if there is an object within 8\arcsec\ with $r_{\rm source}-r_{\rm neighbor} \geq -1$\\
RR2				 		& int32				& 				& Flag if there is an object within 8\arcsec\ with $r_{\rm source}-r_{\rm neighbor} \geq -2$\\
RR25				 	& int32				& 				& Flag if there is an object within 8\arcsec\ with $r_{\rm source}-r_{\rm neighbor} \geq -2.5$\\
RR3				 		& int32				& 				& Flag if there is an object within 8\arcsec\ with $r_{\rm source}-r_{\rm neighbor} \geq -3$\\
RR4				 		& int32				& 				& Flag if there is an object within 8\arcsec\ with $r_{\rm source}-r_{\rm neighbor} \geq -4$\\
RR5				 		& int32				& 				& Flag if there is an object within 8\arcsec\ with $r_{\rm source}-r_{\rm neighbor} \geq -5$\\
PMRA			 		& float32				& \masyear		& Proper motion in R.A. (in proper units, i.e. $\mu_\alpha \cos \delta$)\\
PMDEC			 		& float32				& \masyear		& Proper motion in Decl.\\
PMRA\_M04			 	& float32				& \masyear		& M04 Proper motion in R.A. (in proper units, i.e. $\mu_\alpha \cos \delta$)\\
PMDEC\_M04			 	& float32				& \masyear		& M04 Proper motion in Decl.\\
PMRA\_INTERR			& float32				& \masyear		& Intrinsic error in proper motion in R.A.\\
PMDEC\_INTERR			& float32				& \masyear		& Intrinsic error in proper motion in Decl.\\
PMRA\_MEASERR			& float32				& \masyear		& Measurement error in proper motion in R.A.\\
PMDEC\_MEASERR		& float32				& \masyear		& Measurement error in proper motion in Decl.\\
PMRA\_FITERR			& float32				& \masyear		& Fit error in proper motion in R.A.\\
PMDEC\_FITERR			& float32				& \masyear		& Fit error in proper motion in Decl.\\
PMRA\_TOTERR			& float32				& \masyear		& Combined error in proper motion in R.A.\\
PMDEC\_TOTERR			& float32				& \masyear		& Combined error in proper motion in Decl.\\
PMRAERR\_M04			& float32				& \masyear		& M04 error in proper motion in R.A.\\
PMDECERR\_M04			& float32				& \masyear		& M04 error in proper motion in Decl.\\
BASELINE				& float32				& years			& Time baseline used to compute our proper motions\\
DBIT						& 3 character string		& 				& Detection bit identifying surveys used in computing proper motions\tablenotemark{b}\\
RECOMP					& int32				& 				& Flag indicating proper motions were recomputed (see Section~\ref{neighbors})\\
USE						& int32				& 				& Flag indicating which PM measurement to use\tablenotemark{c}\\
MATCH\_M04				& int32				& 				& Number of SDSS objects within a 1\arcsec\ radius matching the USNO-B object\\
SIGRA\_M04				& float32				& mas			& M04 RMS residual for the proper motion fit in R.A.\\
SIGDEC\_M04				& float32				& mas			& M04 RMS residual for the proper motion fit in Decl.\\
NFIT\_M04				& int32				& 				& Number of detections used in the M04 fit\\
O\_M04					& float32				& mags			& Recalibrated USNO-B O magnitude, recalibrated to SDSS $g$\\
J\_M04					& float32				& mags			& Recalibrated USNO-B J magnitude, recalibrated to SDSS $g$\\
WS\_DIST					& float32				& arcsec			& Total distance between \textit{WISE} position and SDSS position\tablenotemark{d}\\
S2\_DIST					& float32				& arcsec			& Total distance between SDSS position and 2MASS position \tablenotemark{d}\\
W2\_DIST					& float32				& arcsec			& Total distance between \textit{WISE} position and 2MASS position\tablenotemark{d}
\enddata
\tablenotetext{a}{Defined as $.5 \times $(W1MJDMAX-W1MJDMIN).}
\tablenotetext{b}{`111': \textit{WISE}, SDSS, and 2MASS were used; `110': SDSS and 2MASS were used; `011': \textit{WISE} and SDSS were used; `000': SDSS+USNO-B measurement is available.}
\tablenotetext{c}{`1': proper motions were measured here; `2': proper motions are from M04; or `3': both proper motions are available.}
\tablenotetext{d}{Total distance $= \sqrt{(\Delta \alpha)^2 \cos \delta_1 \cos \delta_2 + (\Delta \delta)^2}$.}
\end{deluxetable*}

\bibliography{arxiv}
\bibliographystyle{apj}

\end{document}